\documentclass[]{raa}                  
\usepackage{graphicx,times}             
\input{epsf.sty}                        
\input{psfig.sty}                       

\begin{document}

   \title{Relativistic reflection X-ray spectra of accretion disks  
}

   \volnopage{Vol.0 (200x) No.0, 000--000}      
   \setcounter{page}{1}          

   \author{Khee-Gan Lee
      \inst{1}
   \and Kinwah Wu 
      \inst{2}
   \and Steven V. Fuerst 
      \inst{3} 
      \and Graziella Branduardi-Raymont
      \inst{2} 
   \and Oliver Crowley 
      \inst{2,4} 
   }

   \institute{ 
     Department of Astrophysical Sciences, Princeton University, Princeton, NJ 08544, USA \\
        \and
              Mullard Space Science Laboratory, University College London, Holmbury St.\ Mary,  
      Surrey RH5 6NT, United Kingdom 
      \\
        \and
            Kavli Institute for Particle Astrophysics and Cosmology, Stanford University, Stanford, 
    CA 94309, USA  \\ 
               \and  
             Department of Physics \& Astronomy, University College London, Gower Street, 
     London WC1E 6BT, United Kingdom   \\ 
   }

   \date{Received~~2009 month day; accepted~~2009~~month day}

\abstract{ 
We have calculated the relativistic reflection component of the X-ray spectra of accretion disks 
     in active galactic nuclei (AGN).   
Our calculations have shown that the spectra can be significantly modified 
    by the motion of the accretion flow 
    and the gravity and rotation of the central black hole.  
The absorption edges in the spectra 
   suffer severe energy shifts and smearing, and  
   the degree of  distortion depends on the system parameters, 
   in particular, the inner radius of the accretion disk 
    and the disk viewing inclination angles.  
The effects are significant. 
Fluorescent X-ray emission lines from the inner accretion disk 
   could be powerful diagnostic of space-time distortion and dynamical relativistic effects 
   near the event horizons of accreting black holes.  
However, improper treatment of the reflection component in fitting the X-ray continuum  
   could give rise to spurious line-like features.   
These features mimic the true fluorescent emission lines and may mask their relativistic signatures.  
Fully relativistic models for reflection continua together with the emission lines are needed 
  in order to extract black-hole parameters from the AGN X-ray spectra.   
}

   \authorrunning{K.-G. Lee et al.}            
   \titlerunning{Relativistic reflection X-ray spectra of accretion disks}  

   \maketitle

%
%
\section{Introduction}           
\label{sect:intro}

The power-law component in the keV X-ray spectra of active galactic nuclei (AGN) 
    is believed to be produced by inverse-Compton scattering of low-energy photons  
    from the accretion disk by energetic electrons in a hot corona above the disk 
    (Thorne \& Price 1975; Sunyaev \& Titarchuk 1980; Pozdnyakov, Sobel \& Sunyaev 1983).      
The scattered emission is nearly isotropic.    
A fraction of it emerges directly as the observed power-law component;  
    the rest of it is incident onto the accretion disk, 
    where it is further reprocessed by the cooler disk atmosphere. 
The back irradiation leads to the formation of fluorescent lines   
   (most prominently, the neutral Fe K${\alpha}$ line at 6.4 keV) and 
   a Compton-reflection continuum (for an overview, see Reynolds \& Nowak 2003).  
These fluorescent lines have small thermal widths. 
Although they are intrinsic narrow, 
    they can be severely broadened by scattering, relativistic motions of accretion-disk flows   
    and strong gravity of the black hole. 
Their profiles can also be modified by line-of-sight materials through absorption and scattering.  
The fluorescent lines are therefore powerful diagnostic of the inner AGN environments.  
They have been studied intensively in the last few decades 
   (e.g., Cunningham 1976; Fabian et al.\ 1989; Stella 1990; Laor 1991; Fantan et al.\ 1997; 
   Dabrowski et al.\ 1997; Pariev \& Bromley 1998; Reynolds et al.\ 1999; Beckwith \& Done 2004; 
   Fuerst \& Wu 2004, 2007; Wu et al.\ 2006, 2008; Wu, Ball \& Fuerst 2008).  
The Compton-reflection continuum, like the lines, can be modified 
    by relativistic motion of the flows, gravity of the black hole, and line-of-sight absorption and scattering,  
    thus it is also a useful diagnostic of AGN accretion. 
In spite of this, the relativistic reflection continuum component in the AGN spectrum  
   has not been subject to a similar degree of attention as the fluorescent lines.     

We conduct a systematical analysis 
    of the effects of gravity and relativistic motion on the Compton-reflection spectra of AGN. 
We quantify the distortion of the spectral features 
    and investigate their dependence on various system parameters:  
    the inner and outer radii of the accretion disk, viewing inclination angle, 
    and spin of the black hole.        
We organize the paper as follows.  
 \S2 describes the formulation and the numerical algorithm that we use in the spectral calculations. 
 \S3 presents reflection spectra in various settings  
      and illustrates how the spectral features vary with system parameters.  
\S4 discusses the astrophysical implications and applications. 

\section{Spectral calculations}
\label{sect:spect_calc}

We consider azimuthally symmetric, geometrically thin accretion disks in Keplerian rotation.    
The reflection emission originates from the disk surface layer.  
It is anisotropic and is subject to limb effects. 
We ignore any line-of-sight emission and extinction not directly related to the accretion disk.     
The photons are gravitationally lensed and their energies are relativistically shifted.   
The photon trajectories (the light rays) are determined 
    by the geodesic equation for the space-time specified by the black hole.  
The total spectrum of the accretion disk 
    is the direct sum of contributions of all emitting disk surface elements. 
The spectral calculations consist of three parts: 
   (1) generation of local reflection rest-frame spectra at different viewing angles,  
   (2) calculation of local energy shifts and pitching angles of emerging photons 
         that can reach the distant observer 
         for each surface element on the accretion disk, 
   and (3) convolution of the reflection rest-frame spectra,  
        and the energy shifts and pitching angles of the photons 
        for all disk surface elements.        

\subsection{Reflection rest-frame spectra} 

\begin{figure}
\begin{center} 
\vspace*{0.2cm} 
  \mbox{\epsfxsize=0.7\textwidth\epsfbox{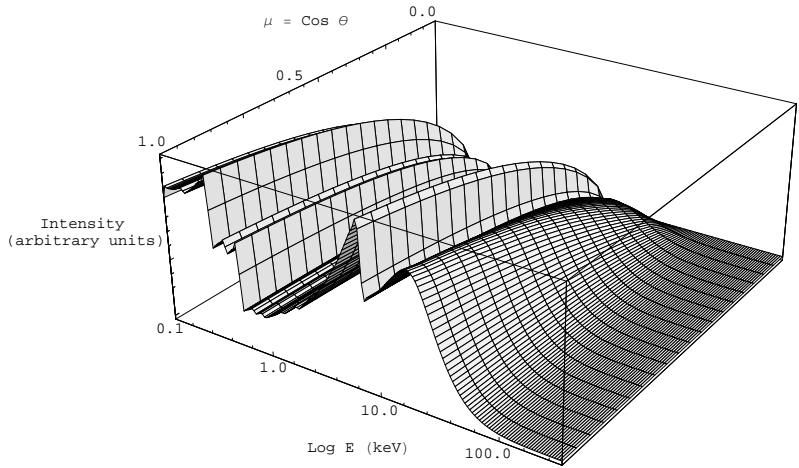}}   
\vspace*{-0.25cm}  
\end{center}
\caption {A surface plot showing the angle-dependence of reflection rest-frame spectra,  
       generated by  {\scshape pexrav}  in {\scshape xspec}. 
   The photon pitching angle $\theta$ is specified by the variable $\mu$ ($\equiv \cos \theta$).  
   The metal abundance is set to be solar. 
   All other parameters are fixed. 
   The intensity is on a linear scale and in arbitrary units. 
   The carbon (C), Oxygen (O) and Iron (Fe) edges (at 0.29, 0.54 and 7.1~keV respectvely) 
        are clearly visible.}
\label{3dplot}
\end{figure}

The main features in the reflection spectra are the continuum and edges. 
The most prominent edges are  
   the carbon (C)  edge at 0.288 keV, oxygen (O) edge at 0.538 keV 
   and iron (Fe) edge at 7.117 keV.  
The reflection emission is anisotropic.    
The depths of edges and the strength of the underlying continuum 
   all vary with the viewing inclination angle.     
We consider the prescription given in Magdziarz \& Zdziarski (1995) (hereafter MZ95)  
   to model the local reflection rest-frame spectra.  
The MZ95 spectra are closed-form approximations to Monte Carlo calculations 
   of the reflected photons' Green's functions.  
They are generated using the {\scshape pexrav} routine 
   in the {\scshape xspec} package (Shafer, Haberl \& Arnaud 1991).     
Figure~\ref{3dplot} shows the reflection rest-frame spectra for  a full range of $\mu$  
   ($\equiv \cos \theta$, where $\theta$ is the pitching angle of the photons). 
The variations of the edges and the continuum with $\mu$ are clearly visible.      

\subsection{Energy shift and lensing of the disk emission}  

\begin{figure}[t] 
\begin{center} 
\vspace*{0.25cm} 
  \mbox{\epsfxsize=0.35\textwidth\epsfbox{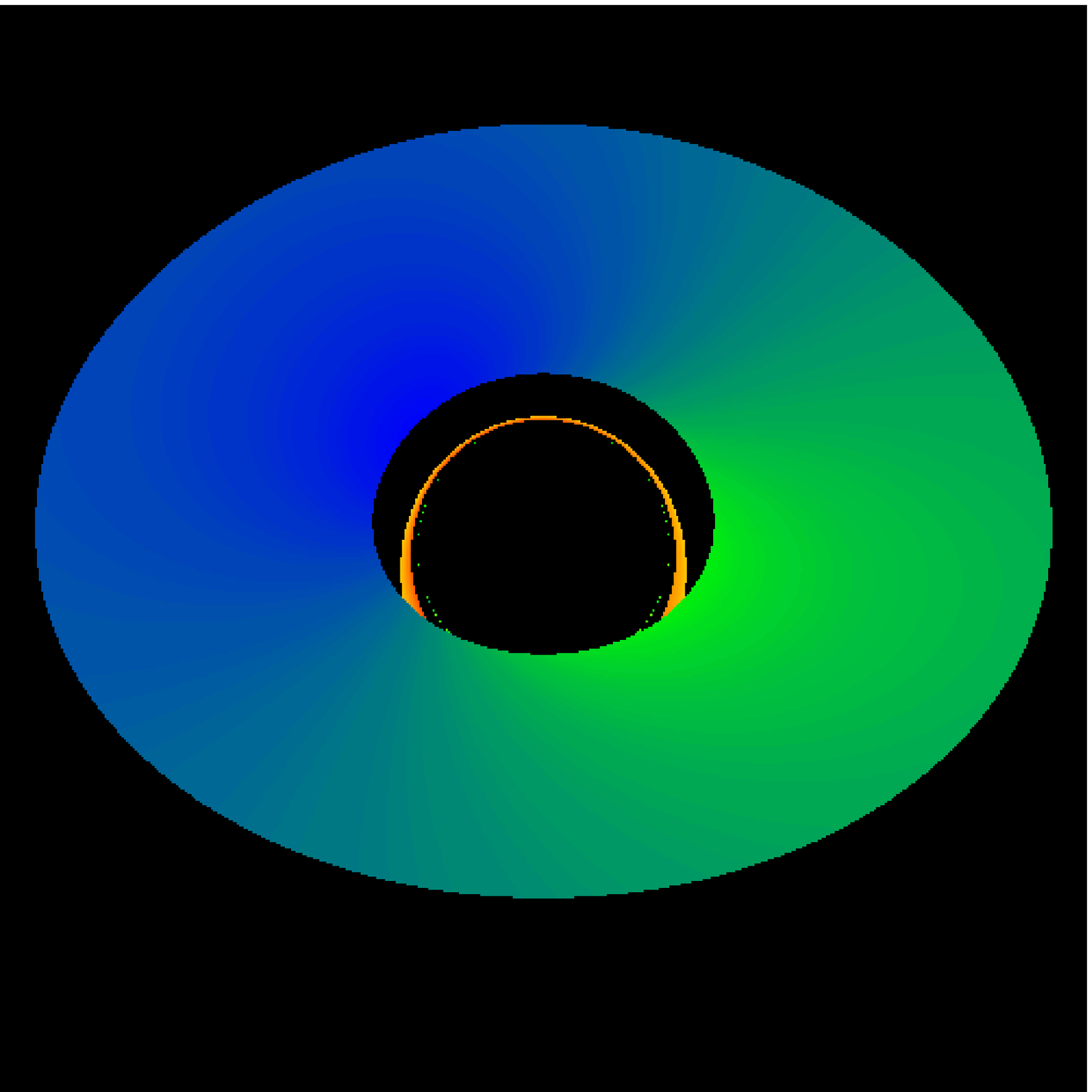}}  \hspace*{0.96cm}
  \mbox{\epsfxsize=0.35\textwidth\epsfbox{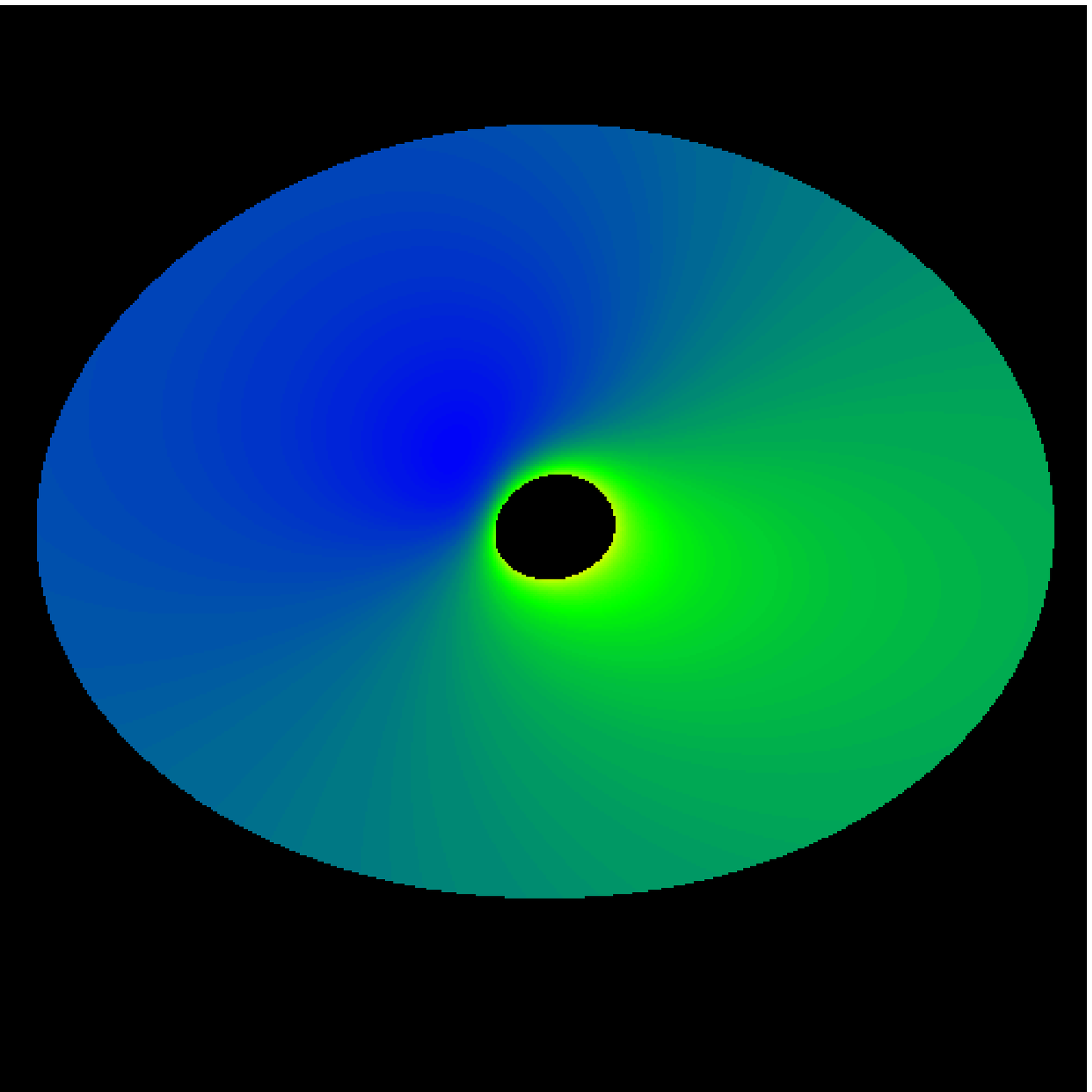}}  \\ 
  \mbox{\epsfxsize=0.35\textwidth\epsfbox{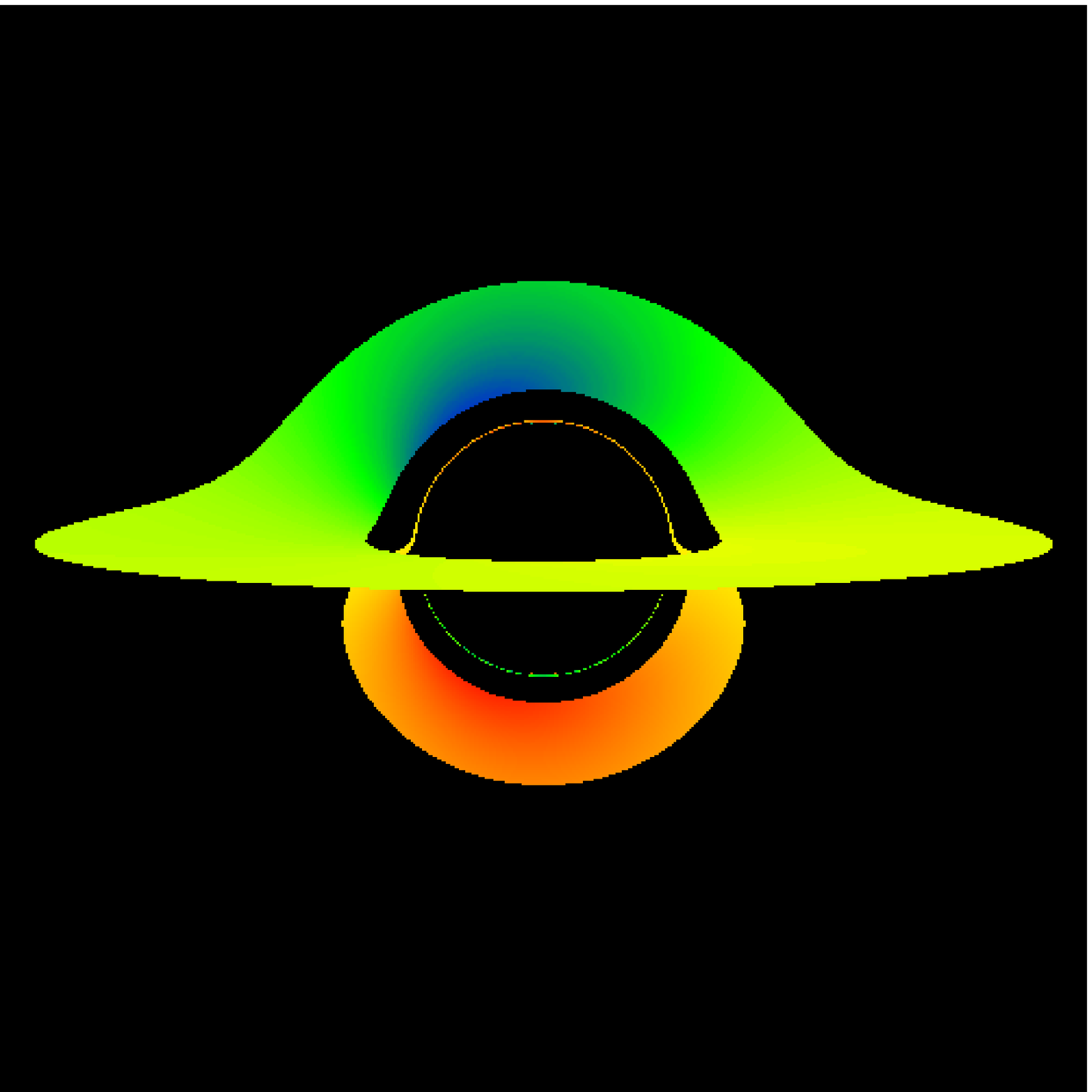}}  \hspace*{0.96cm}
  \mbox{\epsfxsize=0.35\textwidth\epsfbox{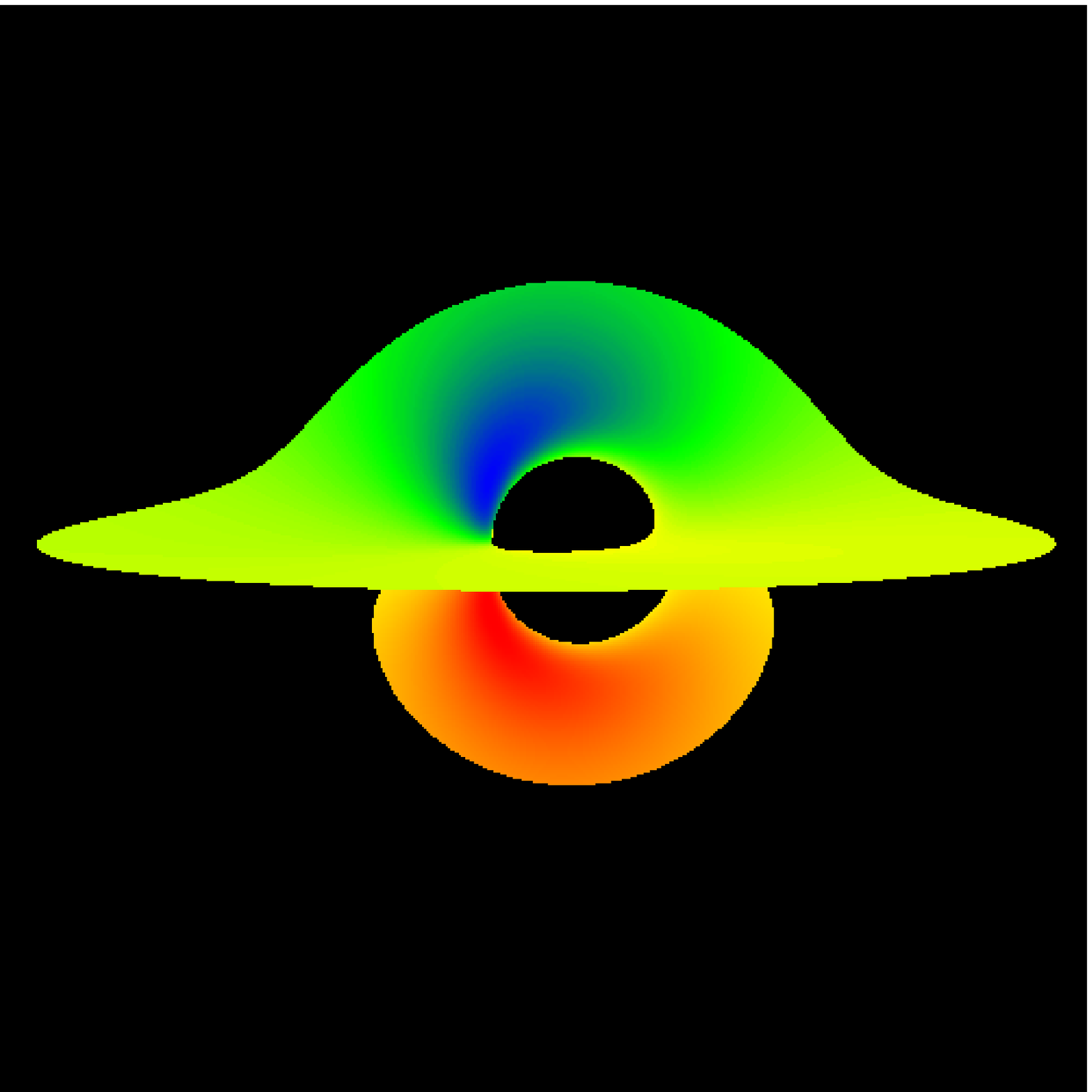}}  \\ 
  \mbox{\epsfxsize=0.42\textwidth\epsfbox{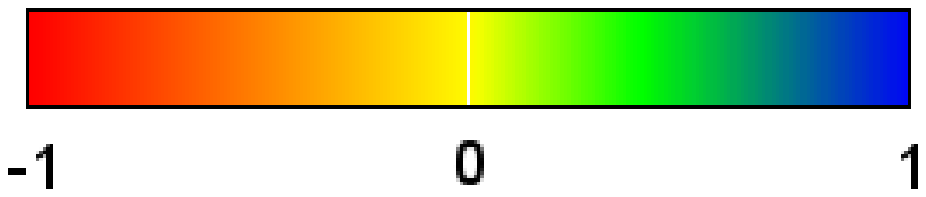}} 
  \mbox{\epsfxsize=0.42\textwidth\epsfbox{ms78fig02x.eps}} 
\end{center}
\vspace*{-0.75cm}
\caption{The distributions of  the pitch angles $\mu$ of the photons,  
        which can reach the observer,   
        on accretion disks 
        around  a Schwarzschild black hole (left column) 
        and a Kerr black hole with $a=0.998$ (right column), 
        for viewing inclinations angles of $45^\circ$ (top panels) and $85^\circ$ (bottom panels). 
 For the Schwarzschild black hole, 
        the inner disk radius is $6~r_{\rm g}$ and the outer disk radius is $20~r_{\rm g}$, 
        where $r_{\rm g} = GM/c^2$ is the gravitational radius, 
        $G$ is the gravitational constant, $c$ is the speed of light, and 
        $M$ is the black-hole mass.  
 For the Kerr black hole,  
     the inner disk radius is $1.23~r_{\rm g}$, and the outer disk radius is $20~r_{\rm g}$.       
 The colour code bars indicate the value of $\mu$, 
        with positive values for  photons emitted from the `top' surface 
        and negative valuses for photons emitted from the `bottom' surface.    
   }
\label{anglepic0}
\end{figure}   

The energy shifts and the degree of boosting of the emission depend on 
   the flow speeds of the material in the emitting surface element relative to the observer 
   and on the gravitational potential difference between the emitter and the observer.   
The energy-shift factor is given by $g = E_{\rm o} / E_{\rm e}$,   
   where $E_{\rm o}$ and $E_{\rm e}$ are the observed and emitted energies
   of the photons respectively.         
The photon flux at a local rest-frame on a disk surface element and the photon flux 
  measured by the observer differ by a multiplication factor of $g^4$, 
   where the factor of $g^3$ takes account of intensity boost due to energy (frequency) shift,   
  and an additional factor of $g$ takes account of time dilation.  

For each surface element photons of a specific pitching angle can reach the distant observer, 
    and other photons emitted from the element will not be seen by the observer.   
Because of gravitational lensing, the pitching angles of these photons 
   are not the same as the viewing inclination angle of the accretion disk.   
They are also affected strongly by the local flow velocities 
   and the rotation frame-dragging of the black hole. 
Combining all these effects makes the pitching angle of the visible photons   
   vary very significantly across the accretion disk.   

We consider a backward ray-tracing algorithm and use the formulation given in Fuerst \& Wu (2004) 
    for the radiative transfer calculations. 
We first determine the energy shift factor $g$ and  the cosine of the patch angle $\mu$
     of the photons for each surface element in the accretion disk
     and then generate disk images (projected in the sky-plane).  
The outputs of calculations for each set of accretion-disk and black-hole parameters 
   are stored and will be  used as inputs in the convolution calculations.   

Figures~\ref{anglepic0} 
   show distributions of the photon pitching angles in accretion disks  
   around a Schwarzschild black hole and a Kerr black hole 
   with spin parameter $a=0.998$ respectively.      
For the disk around a Schwarzschild black hole viewed at $45^\circ$, 
   most photons from the first-order disk image that reach the observer 
   have pitching angles very different from $45^\circ$.  
Moreover, a very substantial fraction of these photons 
   leave the disk at very small pitching angles.         
Even at very high viewing angles (e.g.\ $85^\circ$),  
   the fraction of small pitching-angle photons is non-negligible.  
The variation in the pitching angles of photons across the accretion disk   
   is more obvious for Kerr black holes.      
Photons from accretion disks around Kerr black holes are strongly lensed, 
   and as in the case of the Schwarzschild black hole,   
   the pitching angles of the photons that reach the observer  
   are very different to the disk viewing inclination angle.   
Rotational frame-dragging effects are important.   
As shown in Figure~\ref{anglepic0},  
  rotational frame dragging effects are very noticeable 
  in the inner disk regions  at a moderate disk viewing inclination angle  
  ($45^\circ$)
  as well as at a very high disk viewing inclination angle ($85^\circ$). 

We note that rotational frame dragging has stronger effects  
  on the pitching angles than on the energy shifts of the photons.   
The reflection continuum, which is anisotropic, is thus  
  more sensitive to the black hole's rotation than the lines, 
  which depend on the black-hole spin parameter 
  via a constraint on the inner radius of the accretion disk.    

\subsection{Spectral convolution}  

The spectral convolution calculations 
   require local reflection spectra with a range of photon pitching angles as imput, 
   but  {\scshape xspec} only generates spectra at fixed viewing angles.  
We therefore produce grids of spectra for a range of $\mu$ 
   and interpolate between the grids for any arbitrary value of $\mu$.  
We consider the Neville's algorithm (Press et al.\ 1992)  
  for n-th order polynomial interpolation.   
The key features in the reflection spectra are the edges, 
   otherwise the spectra consist of piecewise smooth curves (see Fig.~\ref{3dplot}).  
As a compromise between the two extremes, 
   we use 5-th order polynomials in the spectral interpolations.  

The emissions from each disk element are associated with a value of $g$ and $\mu$.  
We shift the energy of the binned spectrum of each disk element 
    according to the corresponding $g$ value and rescale the intensity by the factor $g^4$.   
We  then re-bin the spectra, 
  as  the energy grid-points of the shifted spectra are generally not  aligned with each other. 
   
Finally, we specify the weights (relative contribution) of the reflection emission  
   from the disk elements. 
We consider a parametric model for the disk emissivity which takes a  power-law form: 
$I(E_{\rm e}, r_{\rm e}) \propto r_{\rm e}^{-\gamma}$, 
    where $I({E_{\rm e}})$ is the emitted intensity at energy $E_{\rm e}$,  
    $r_{\rm e}$ is the radial distance of the emitting disk element from the central black hole, 
    and $\gamma$ is the power-law index.  
Summing all these weighted `corrected' reflection spectra from the disk surface elements 
   give the total spectrum of  the accretion disk.       

\subsection{Basic parameters and other computational settings}  

\begin{table} 
\begin{center} 
\caption[]{Basic set of system parameters used in the calculations}
\label{tab1}
\begin{tabular}{l c} \hline
black-hole spin parameter, $a$ &\footnotesize 0.0 \\
Inner disk radius, $r_{\rm in}$ &\footnotesize $6\, r_{\rm g}$ \\
Outer disk radius, $r_{\rm out}$ &\footnotesize $20\, r_{\rm g}$ \\
disk inclination angle, $i$ &\footnotesize $ 45^{\circ}$\\
Emissivity power-law index, $\gamma$ &\footnotesize 3\\
Incident photon index, $\Gamma$ &\footnotesize 1.7 \\
Spectral cutoff energy, $E_{\rm c}$ &\footnotesize $150\, {\rm keV}$ \\
Metal abundance (relative to solar) &\footnotesize 1.0 \\
Iron abundance (relative to solar) &\footnotesize 1.0 \\
Spectral  normalisation &\footnotesize 1.0 \\ 
\hline 
\end{tabular} 
\end{center} 
\end{table} 

We define a set of basic parameters (Table~\ref{tab1}).  
Unless otherwise stated, we use this set of parameters in our calculations.  
The resolution of the disk images is chosen to be $250 \times 250$ pixels, 
    sufficient to suppress the numerical noise in the computation.  
The direct disk image and 2 next higher-order lensed images  were included.  
For the grids of reflection rest-frame spectra generated by {\scshape xspec},  
   we set the grid step $\Delta \mu = 0.05$, in the range $0.05 < \mu < 0.95$.   
The energy range of the  spectra is from 0.1~keV to 200~keV.  

\section{Results}
\label{results} 
  
\begin{figure}[t]  
\begin{center} 
\vspace*{0.25cm} 
  \mbox{\epsfxsize=0.5\textwidth\epsfbox{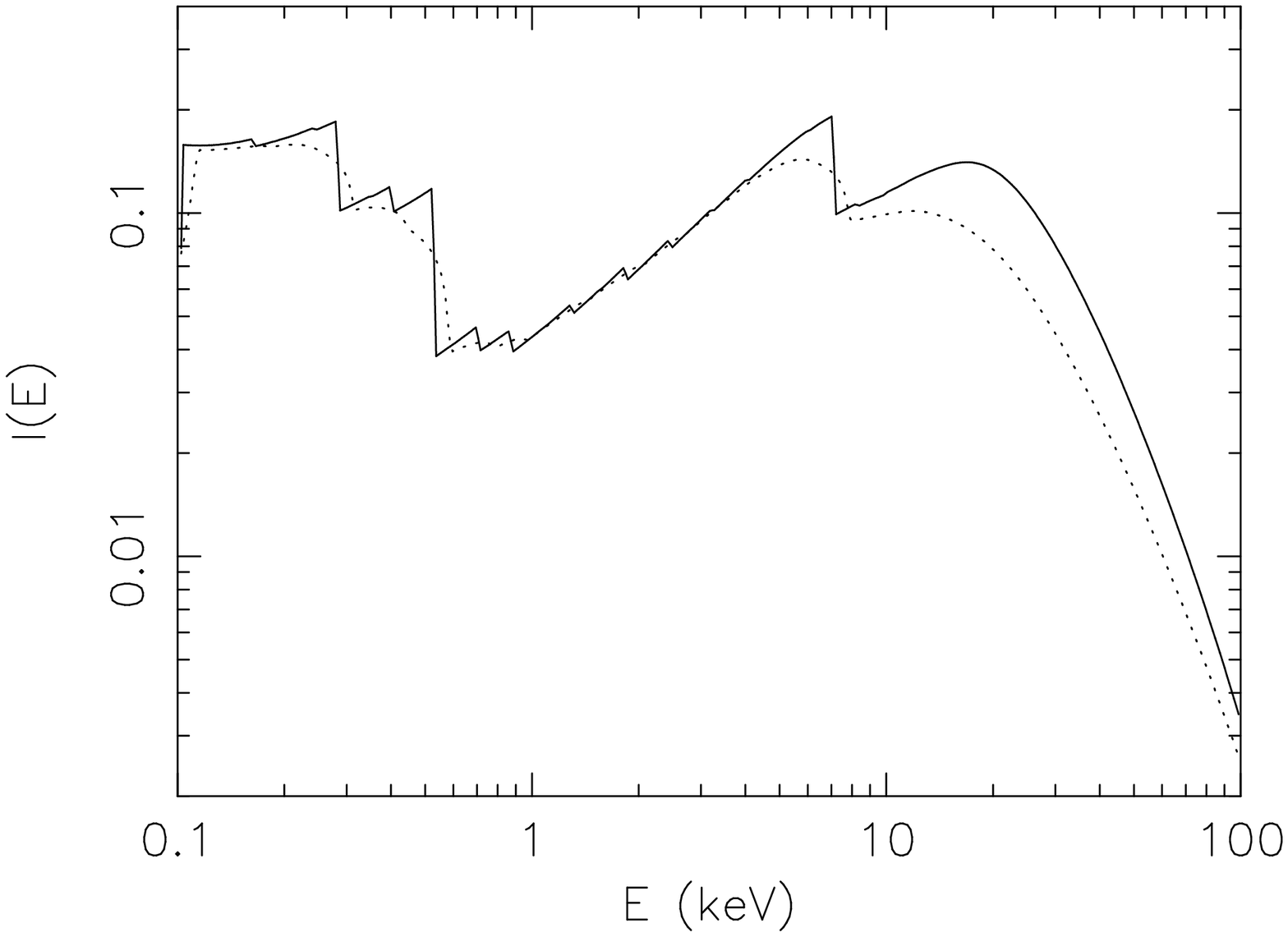}}    \\ 
  \mbox{\epsfxsize=0.5\textwidth\epsfbox{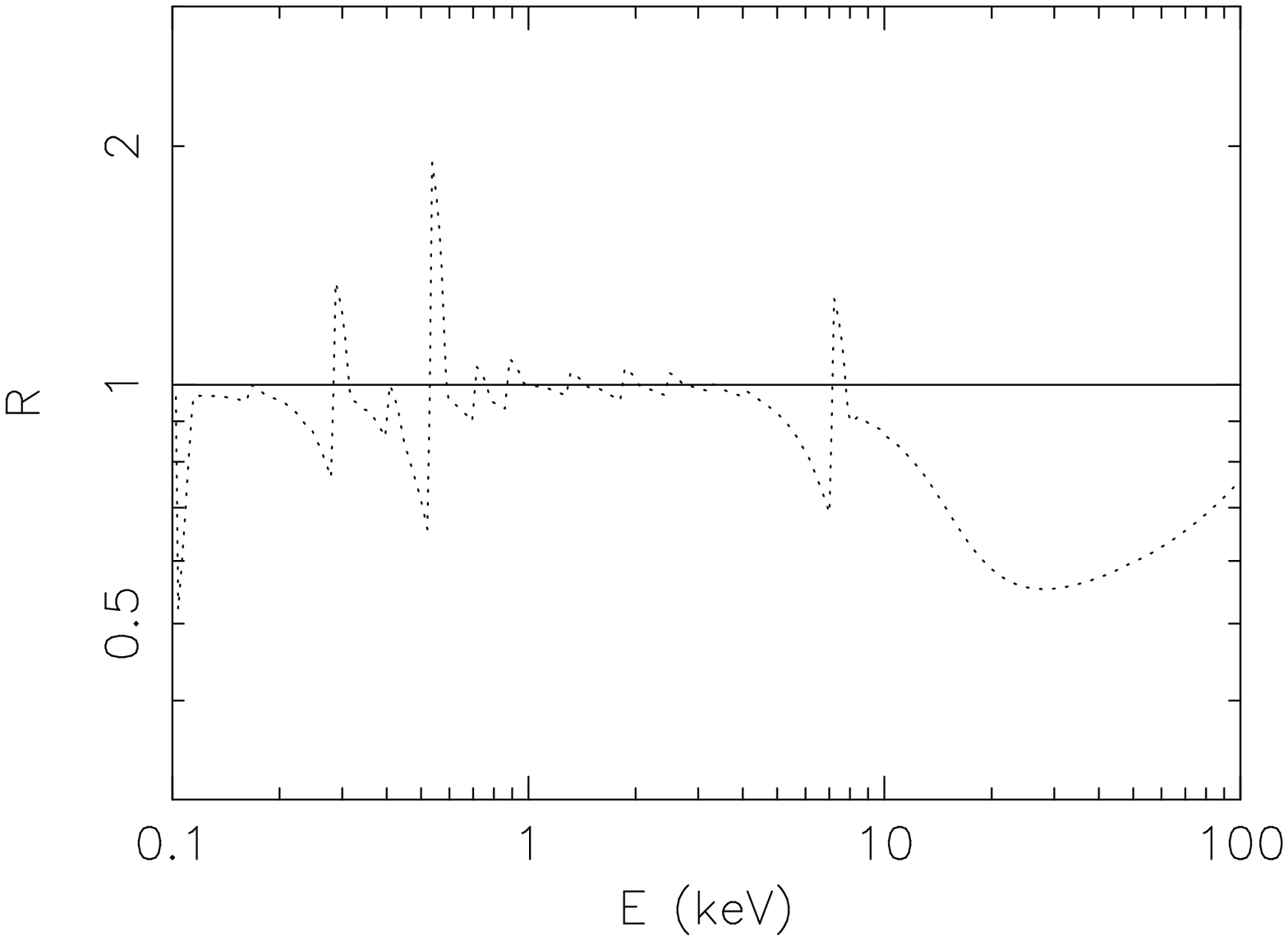}}  
\end{center}
\caption{(Top) 
    A reflection spectrum of a relativistic accretion disk around a Schwarzschild black hole 
        (dotted line) in comparison to a reflection rest-frame MZ95 spectrum 
        at a viewing angle of $\theta = 45^{\circ}$.   
    The intensities of the spectra $I(E)$ are in arbitrary units. 
    (Bottom) The quotient spectrum obtained 
        by dividing the relativistic spectrum by the rest-frame spectrum. 
     The normalisation is such that the normalised intensity $R=1$ at the energy $E = 1$~keV 
         in both cases. }  
\label{shiftplot} 
\end{figure}

The main features in a reflection rest-frame spectrum are the sharp C, O and Fe absorption edges. 
These edges are shifted and smeared in the reflection spectra 
  due to gravitational lensing and various relativistic effects 
   (cf. the spectra in the top panel of Fig.~\ref{shiftplot}, 
   see also the quotient spectrum of the relativistic case in the bottom panel).   
Note that the entire spectrum is  red-shifted, 
   and the effect is more visible at the high-energy ($>$10~keV) part of the spectrum.   
The energy shift is less apparent at the low- and medium-energies 
   due to the profusion of spectral features.  

In the following subsections, 
   we will elaborate and assess 
   how the relativistic reflection spectra are dependent on the system parameters.   
We consider a basic reference case with the parameters listed in Table~\ref{tab1},   
   and unless otherwise stated we vary only one parameter at a time. 
We examined the dependence on the disk inner radius $r_{\rm in}$, 
   disk outer radius $r_{\rm out}$, 
   disk emissivity profile power-law index $\gamma$, 
   black-hole spin parameter $a$ and disk viewing inclination angle $i$.

\subsection{Inner disk radius}

\begin{figure}[t]
\begin{center} 
\vspace*{0.25cm} 
  \mbox{\epsfxsize=0.5\textwidth\epsfbox{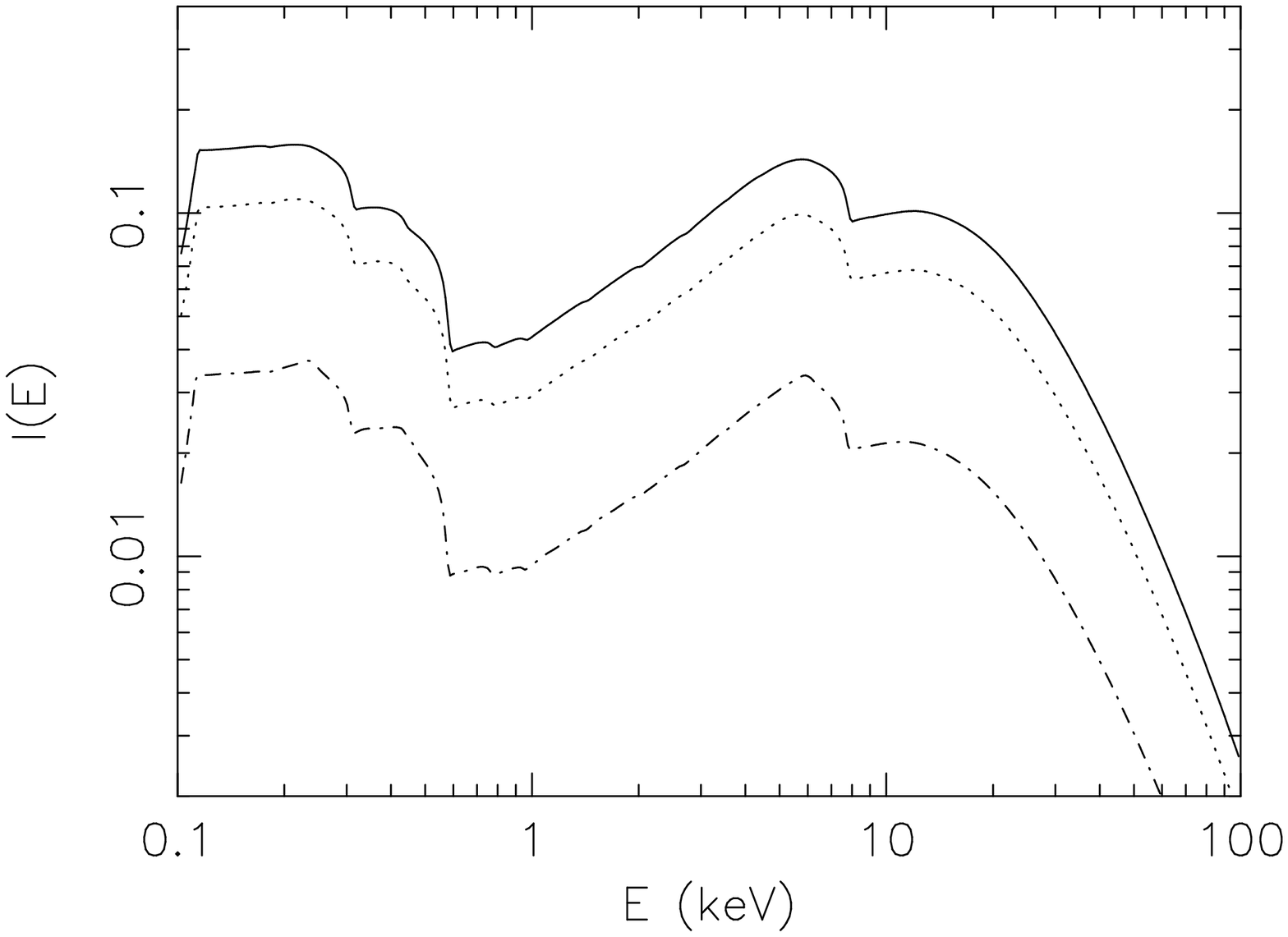}}    \\ 
  \mbox{\epsfxsize=0.5\textwidth\epsfbox{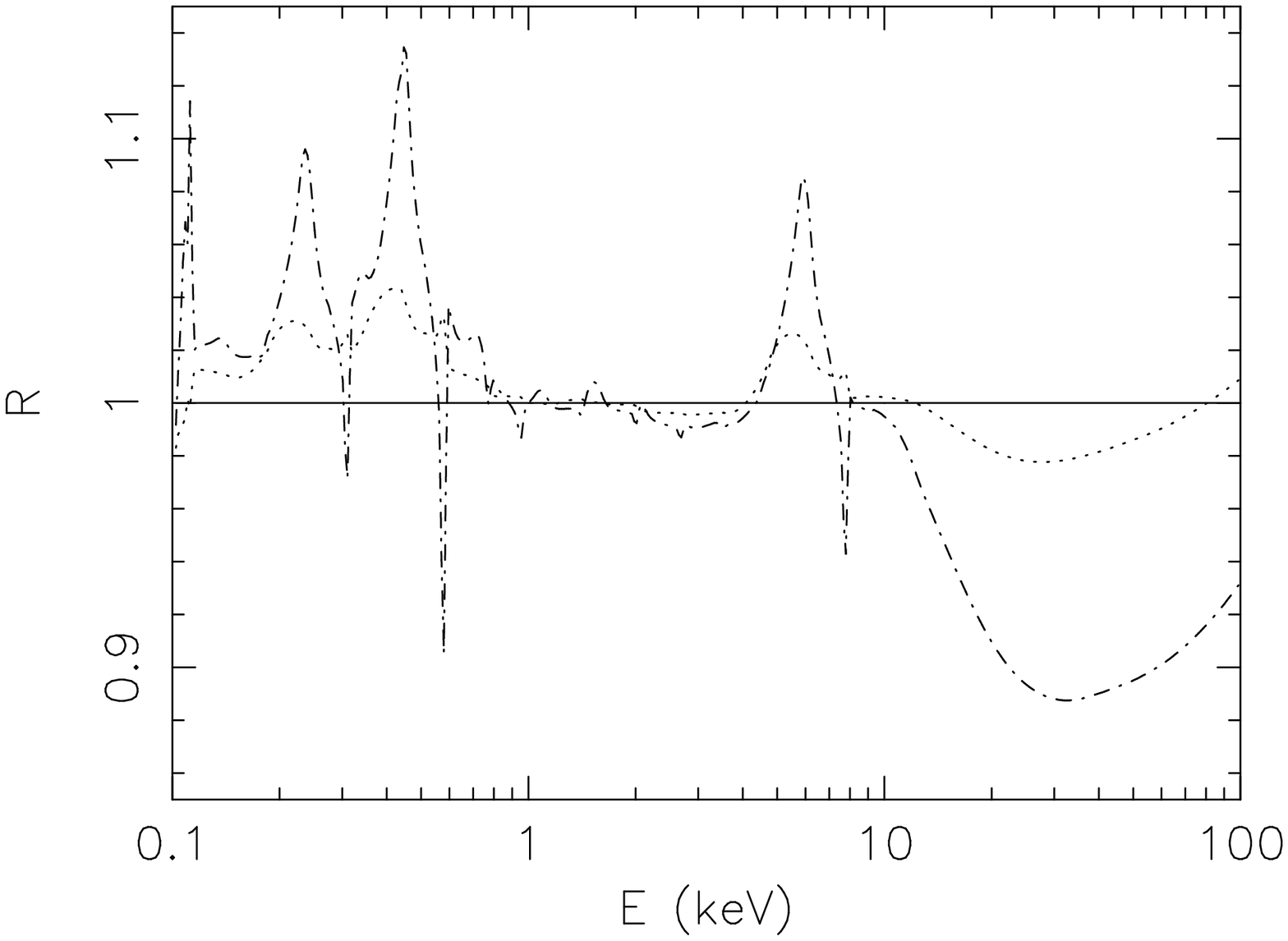}}  
\end{center} 
\caption{ 
   Spectra showing the effects due to changes in the inner radius of the accretion disk, $r_{\rm in}$. 
  The top panel shows the reflection spectra of relativistic accretion disks 
      with $r_{\rm in} = 6$, 10 and 25~$r_{\rm g}$  
      (denoted by solid, dotted and dot-dashed lines respectively).  
  The intensity $I(E)$ is in arbitrary units. 
  The bottom panel shows the corresponding quotient spectra. 
  The reflection spectrum of the case with $r_{\rm in} = 6\ r_{\rm g}$  
      is used as the reference spectrum to derive the quotient spectra. 
   The normalisation is the same as in Fig.~\ref{shiftplot}. } 
\label{rinplot}    
\end{figure}

We begin by investigating the effects of the inner edge $r_{\rm in}$ of the accretion disk. 
We considered disks around non-rotating black holes (with $a=0$).  
The outer disk radius is fixed, with $r_{\rm out} = 100\ r_{\rm g}$  
    but the inner radius is the parameter to vary.  
($r_{\rm g} = G M/c^2$ is the gravitational radius, 
    where $M$ is the black-hole mass and $c$ is the speed of light.) 
Figure~\ref{rinplot} shows spectra for three values of the  inner radius,  
   $r_{\rm in} = 6$,  10 and 25~$r_{\rm g}$,   
   (top panel) and the corresponding quotient spectra (bottom panel).     
As shown,    
    the brightness of the  disk reflection is dependent on the inner disk radius $r_{\rm in}$: 
    the total intensity is larger for smaller  $r_{\rm in}$. 
This can be explained by the fact that,  
    when $r_{\rm out}$ is fixed,   
    the total disk surface area is determined by $r_{\rm in}$. 
The differences between the three spectra are manifested in the quotient spectra.   
The spiky features in the quotient spectra are due to the difference 
    in the smearing and relative energy shifts of the edges, 
The difference between the spectra for $r_{\rm in}=6$ and 10~$r_{\rm g}$ 
    is smaller than the difference between the spectra for  $r_{\rm in}=6$ and 25~$r_{\rm g}$, 
    indicative of importance of the relativistic effects in the inner region of the accretion disk.  
At energies above 10~keV, the quotient spectra show a broad `dip' shape feature, 
   which is a result of competition 
   between energy-shift and smearing as well as lensing of the emission 
   at the high-energy tails of the reflection rest-frame spectra  
   from the disk surface elements.

\begin{table} 
\begin{center} 
\caption[]{Radius of the last  stable particle orbit, $r_{\rm ms}$, as a function 
      of the black-hole spin parameter $a$. } 
\label{tab2}
\begin{tabular}{c @{\hspace{2cm}} c} \hline
   $a$ & $r_{\rm ms}/r_{\rm g}$ \\ \hline
   0.0      & 6.00  \\
   0.5      & 4.23 \\
   0.998 & 1.23\\
\hline
\end{tabular}
\end{center} 
\end{table}

\subsection{Outer disk radius} 

For large values of the power-law index $\gamma$ of the radial emissivity profile,  
    the contribution to the spectrum is dominated by the inner disk region.   
We therefore choose a disk emissivity power-law law with $\gamma = 1$ 
   (instead of 3 as in other cases)  
   so to  increase the relative contributions from the disk elements in the outer disk regions.

\begin{figure}[t]
\begin{center} 
\vspace*{0.25cm} 
  \mbox{\epsfxsize=0.5\textwidth\epsfbox{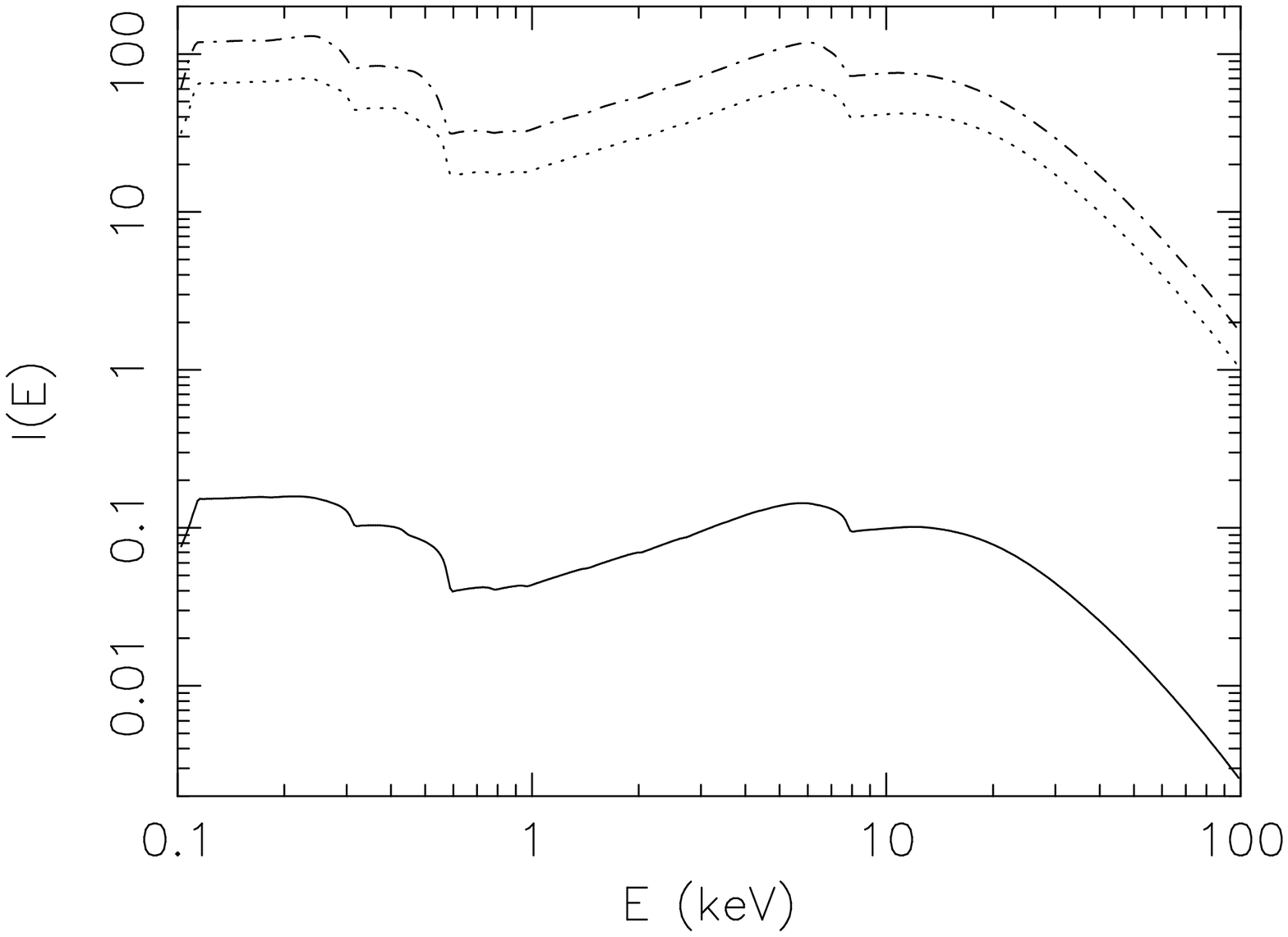}}    \\ 
  \mbox{\epsfxsize=0.5\textwidth\epsfbox{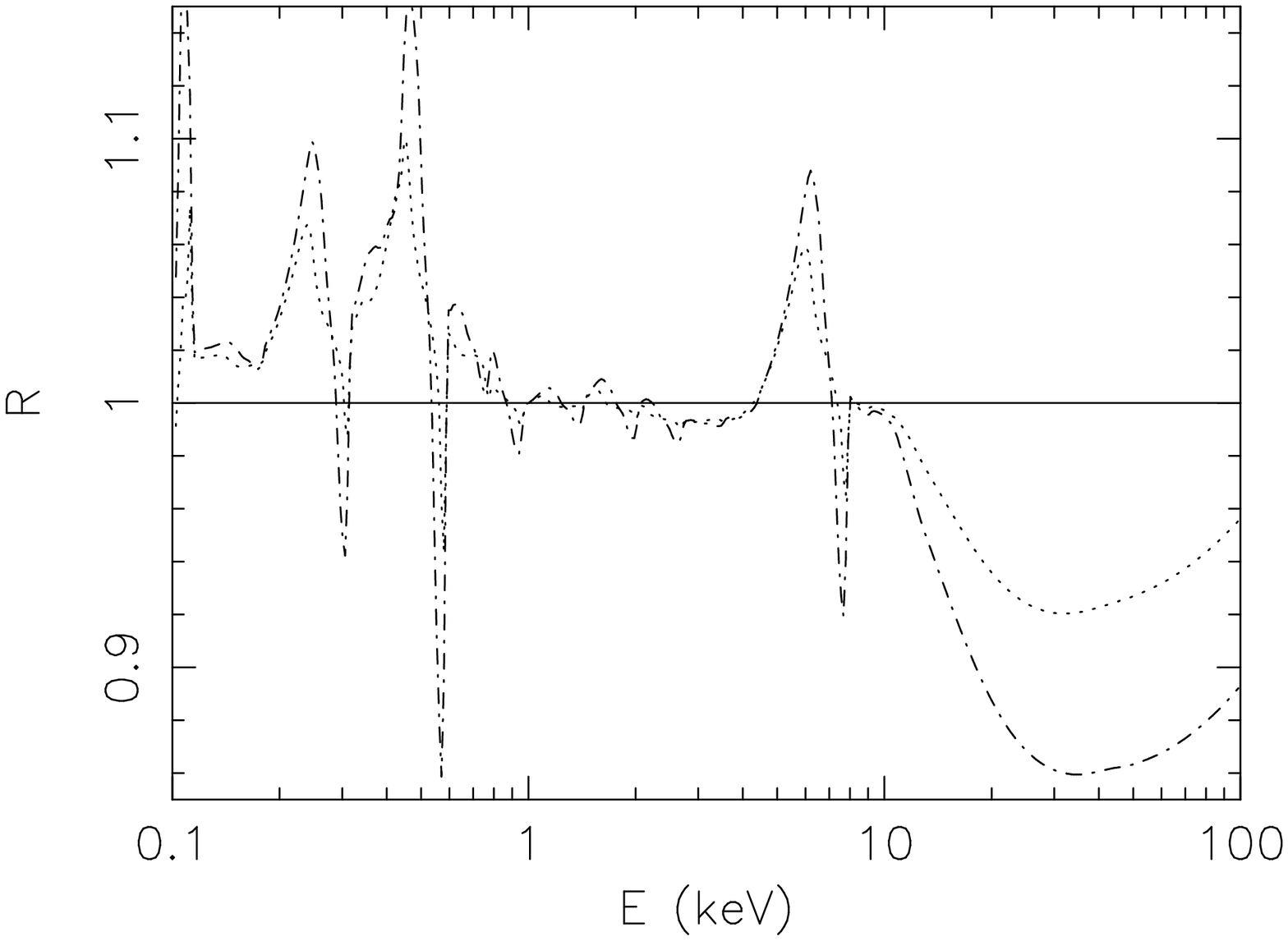}}  
\end{center} 
\end{figure}
\begin{figure}
\caption{Same as  Fig.~\ref{rinplot} but for different disk outer radius $r_{\rm out}$.  
  The reflection spectra and quotient spectra for $r_{\rm out} =$ 20, 50 and 100~$r_{\rm g}$.  
      are denoted by solid, dotted and dot-dashed lines respectively.  
   The emissivity profile is a power law with $\gamma = 1$. 
  The case with $r_{\rm out} =$ 20~$r_{\rm g}$ is used as the reference spectrum 
     to derive the quotient spectra. } 
\label{routplot} 
\end{figure} 

Figure~\ref{routplot} shows three reflection spectra and quotient spectra   
  with $r_{\rm out} = 20$, 50 and 100 $r_{\rm g}$.  
There is less overall emission from the disk with $r_{\rm out}= 20\ r_{\rm g}$, 
  again for the reason that it has a smaller disk surface area.  
There are differences in the three spectra, 
    mostly around the regions of the C, O and Fe features at $\approx$ 0.3, 0.6 and 7~keV.  
The spike and trough features are similar in the quotient spectra, 
     which contrasts with the case of the variable $r_{\rm in}$ 
     (cf.\ Fig.~\ref{rinplot} and Fig.~\ref{routplot}).  
The broad dip feature at energies above 10~keV is clearly visible in the  quotient spectra.  
Here, it shows that adjusting $r_{\rm out}$ 
     changes the relative contribution 
    from the outer disk regions. 
In a large disk with a flat emissivity profile (small $\gamma$), 
    the emission from the outer disk regions 
    tends to overwhelm  the strongly relativistically modified emissions from the inner disk regions. 

\subsection{Disk emissivity power-law index} 

\begin{figure}[t] 
\begin{center} 
\vspace*{0.25cm} 
  \mbox{\epsfxsize=0.5\textwidth\epsfbox{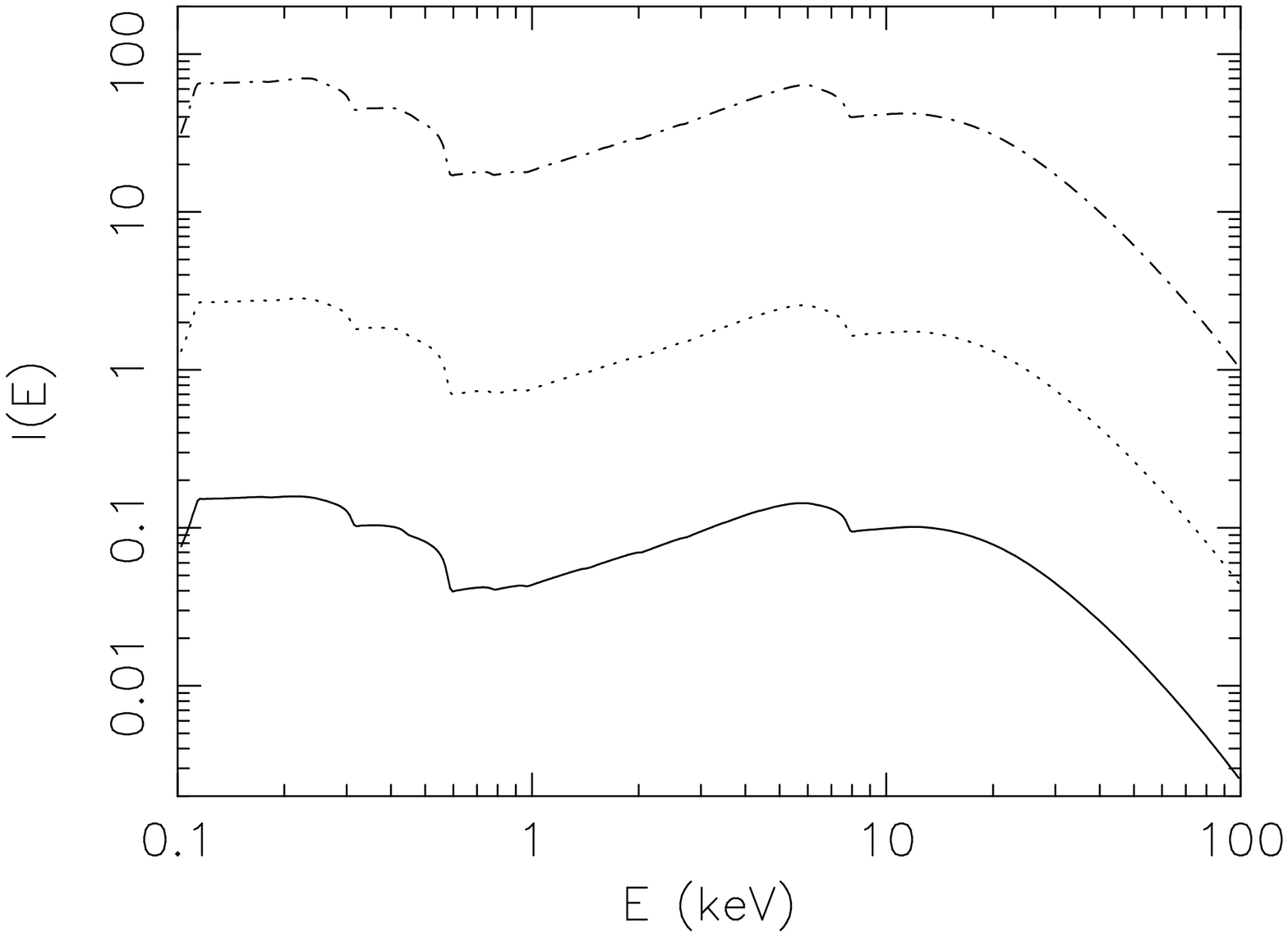}}    \\ 
  \mbox{\epsfxsize=0.5\textwidth\epsfbox{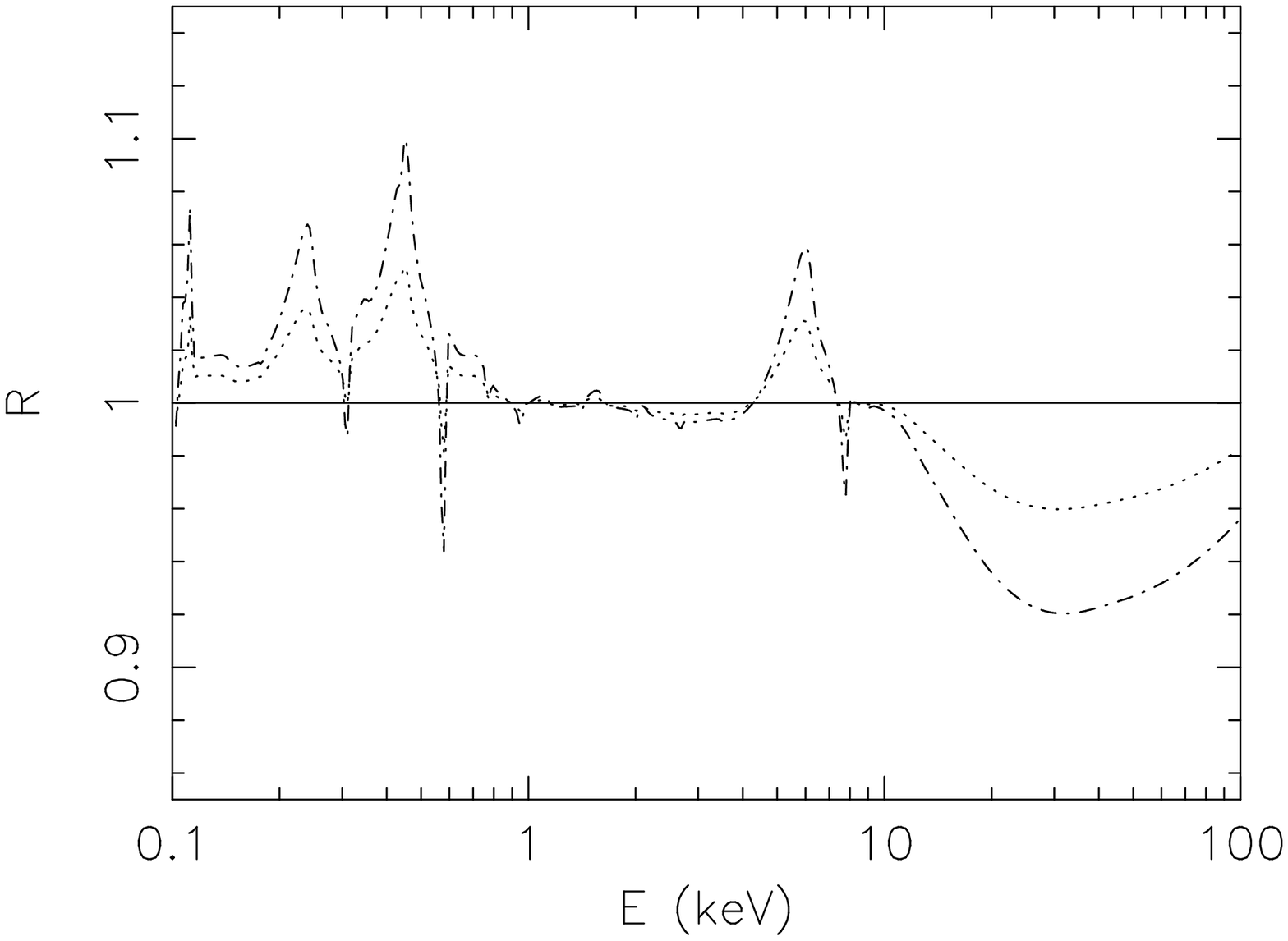}}  
\end{center}
\caption{Same as  Fig.~\ref{rinplot} but for different  disk emission power- law index $\gamma$. 
   The curves in the top panel are the spectra for $\gamma =$ 3, 2, and 1  
      (denoted by solid, dotted and dot-dashed lines respectively);   
      the curves in the bottom panel are the corresponding quotient spectra, 
      with respect to the case with $\gamma =3$.} 
\label{emlawplot}
\end{figure}

The relative contributions of the inner and outer disk regions to the spectrum of an accretion disk 
    are determined by the emissivity power-law index $\gamma$. 
Figure~\ref{emlawplot} shows the spectra and quotient spectra 
    for $\gamma$ = 1, 2 and 3. 
The total intensity of the reflection spectrum decreases as $\gamma$ increases, 
    while the differences between the overall shapes of the spectra  are in the edges. 
The O and Fe edges become slightly less prominent when $\gamma$ decreases, 
   due to the fact that a small $\gamma$ 
   allows more reflection to come from the outer regions of the disk, 
   which are less affected by the relativistic smearing and energy shifting. 
Note that the broad dip feature is present in the quotient spectra. 

\subsection{Spin of the black hole}

Here,  we set the inner rim of an accretion disk  
    equal to the radius of the last stable particle orbit, $r_{\rm ms}$.    
This is essentially to assume of a zero-torque boundary condition at the inner disk rim 
   (Page \& Thorne 1974; Eardley \& Lightman 1975). 
As the last stable particle orbit $r_{\rm ms}$ is determined by the black-hole spin parameter $a$ only,   
    once we have specified $a$,  $r_{\rm ms}$ is also determined.   
Table~\ref{tab2} shows the value of $r_{\rm ms}$ for three cases 
    considered in our calculations (with $a=0$, 0.5 and 0.998, 
    corresponding to a Schwarzschild black hole, a modestly rotating Kerr black hole 
    and a nearly extremal Kerr black hole). 

\begin{figure}[t] 
\begin{center} 
\vspace*{0.25cm} 
  \mbox{\epsfxsize=0.5\textwidth\epsfbox{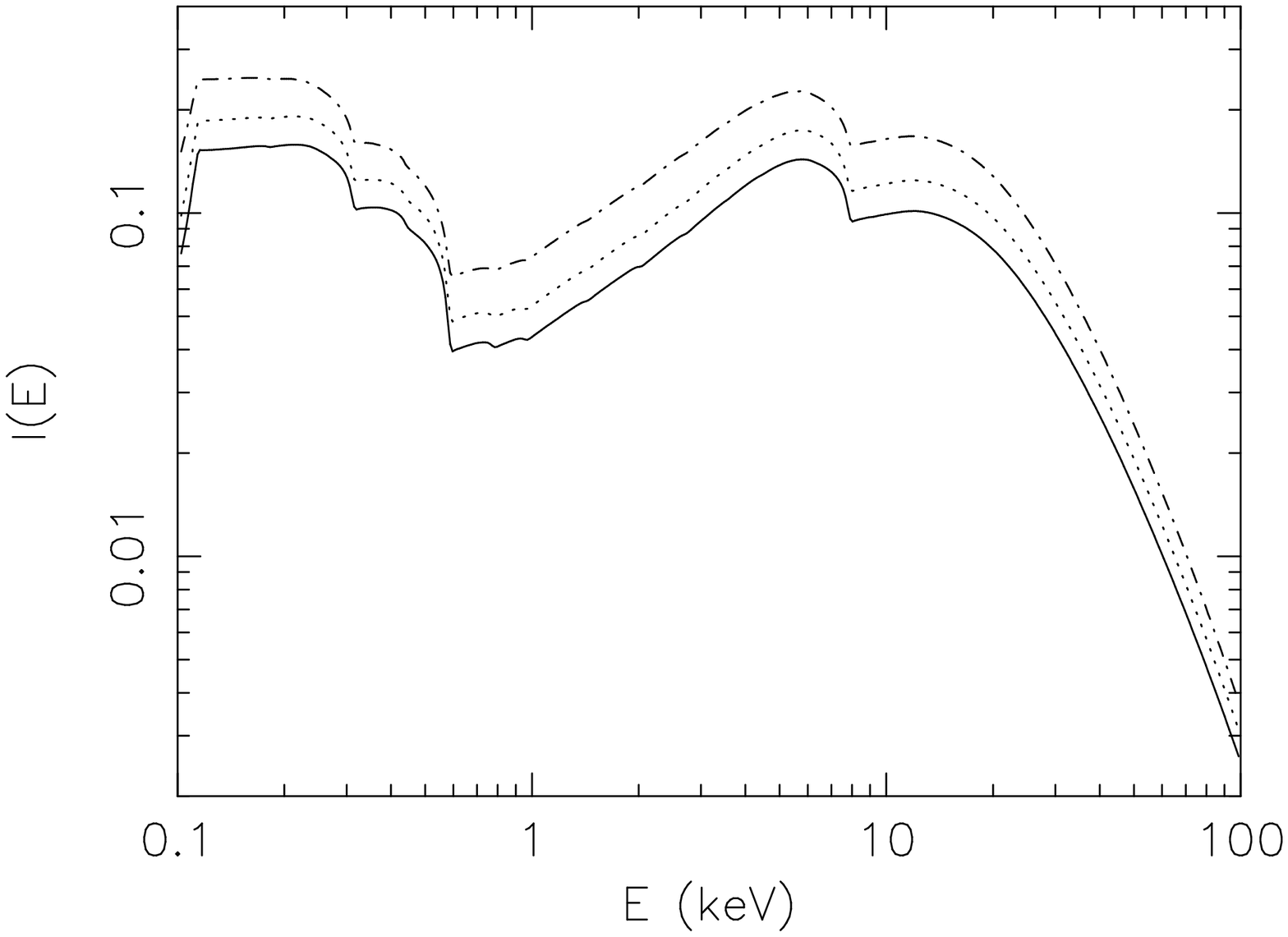}}    \\ 
  \mbox{\epsfxsize=0.5\textwidth\epsfbox{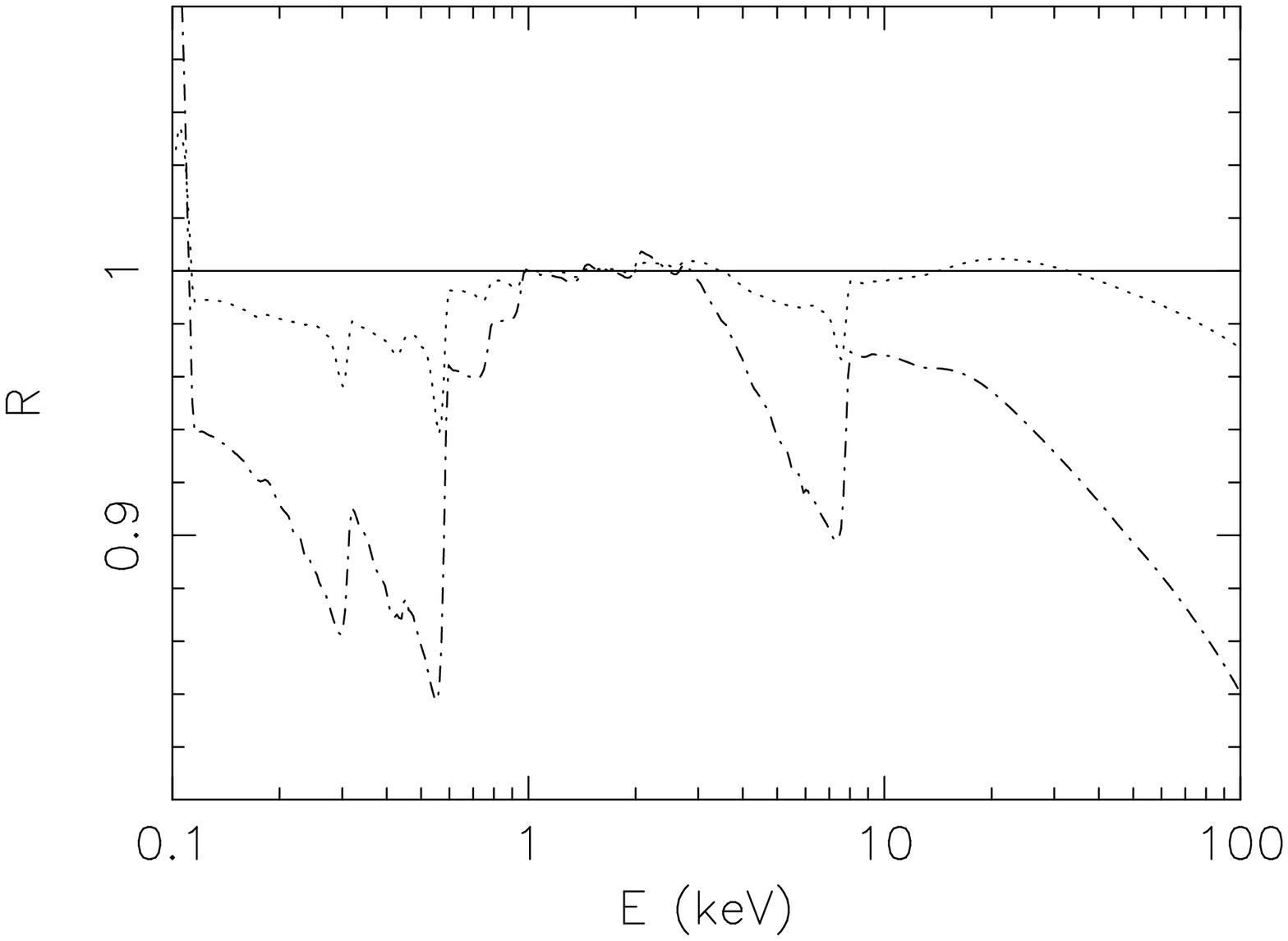}}  
\end{center}
\caption{Same as  Fig.~\ref{rinplot} but for different  values of black-hole spin parameter $a$. 
    Spectra for $a =$ 0, 0.5, and 0.998 (denoted by solid, dotted and dot-dashed lines respectively) 
         are shown in the top panel 
         and the corresponding quotient spectra, relative to the case $a=0$, 
         are shown in the bottom panel.  }
\label{spinplot} 
\end{figure}

The spin of the black hole is an important parameter  
    in determining the accretion disk reflection spectrum.    
An increase in the black-hole spin parameter $a$ 
    increases the total surface area of the accretion disk, if the outer disk radius is fixed. 
An immediate consequence is that this will increase the total intensity of the disk emissions.  
Another consequence is that 
   this will increase the relative weight of the gravitationally shifted and lensed emissions 
   from the innermost disk regions close to the black-hole event horizon,  
   in comparison to the emissions from the outer disk regions.  
In addition, the spin of the black hole determines the degree of rotational frame dragging, 
   which, as shown in \S 2.2, can alter the distribution of pitching angles of the photons, 
   which can be seen by the observer, in the accretion disk.   
In Figure~\ref{spinplot} we show the reflection spectra 
   for the black-hole spin parameters $a=0.0$, 0.5 and 0.998.     
The C, O and Fe edges are shifted to lower energies as $a$ increases.  
The spiky features that had been seen in the quotient spectra previously   
   now become wedge-like,    
   with an extensive low-energy (red) wing and a sharp higher-energy (blue) edge. 
The O and C wedges  overlap substantially in the spectrum of $a=  0.998$. 
The dip feature above 10~keV now disappears from the quotient spectrum. 

\subsection{Disk viewing inclination angle}

\begin{figure}[t] 
\begin{center} 
\vspace*{0.25cm} 
  \mbox{\epsfxsize=0.5\textwidth\epsfbox{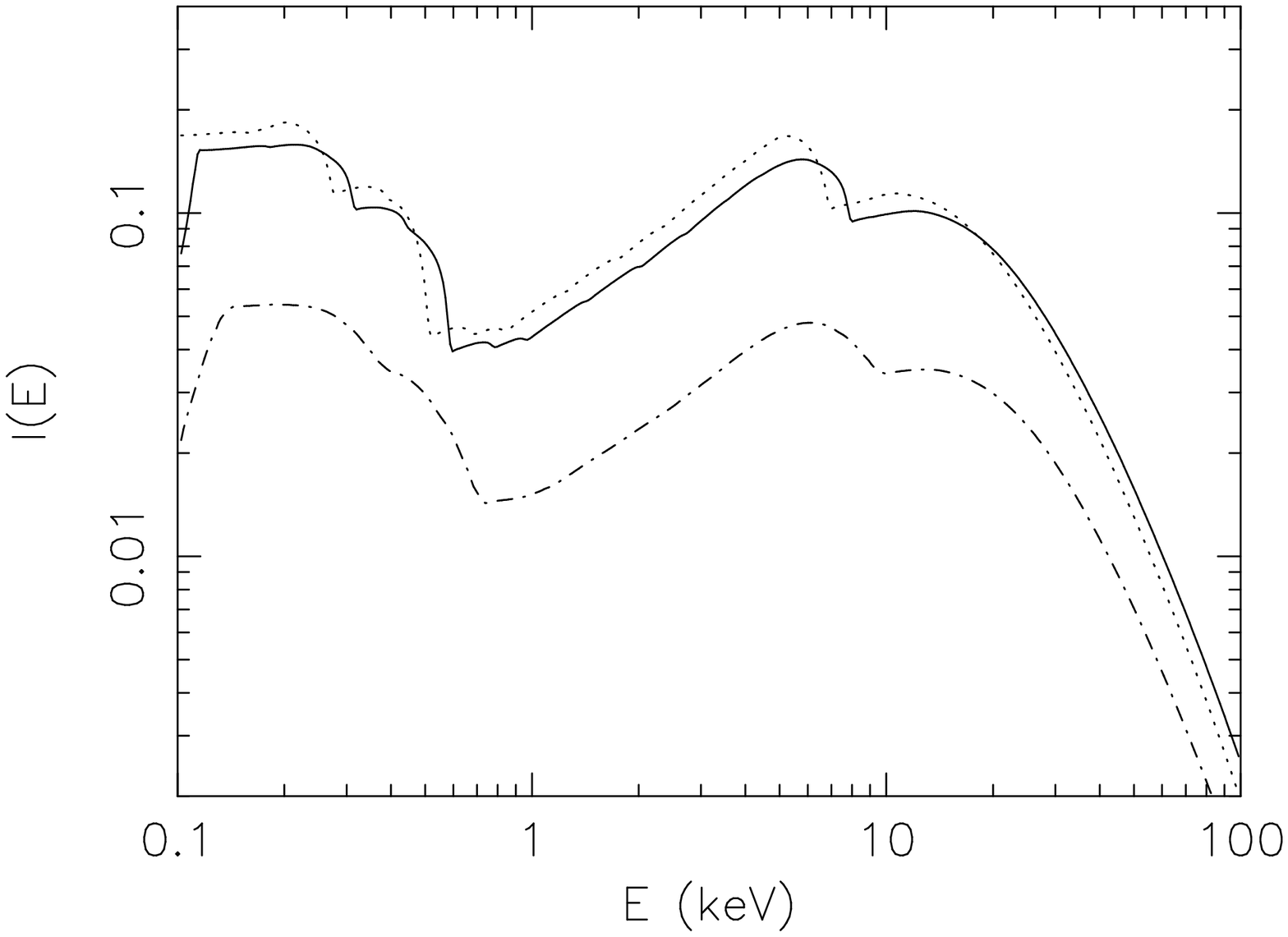}}    \\ 
  \mbox{\epsfxsize=0.5\textwidth\epsfbox{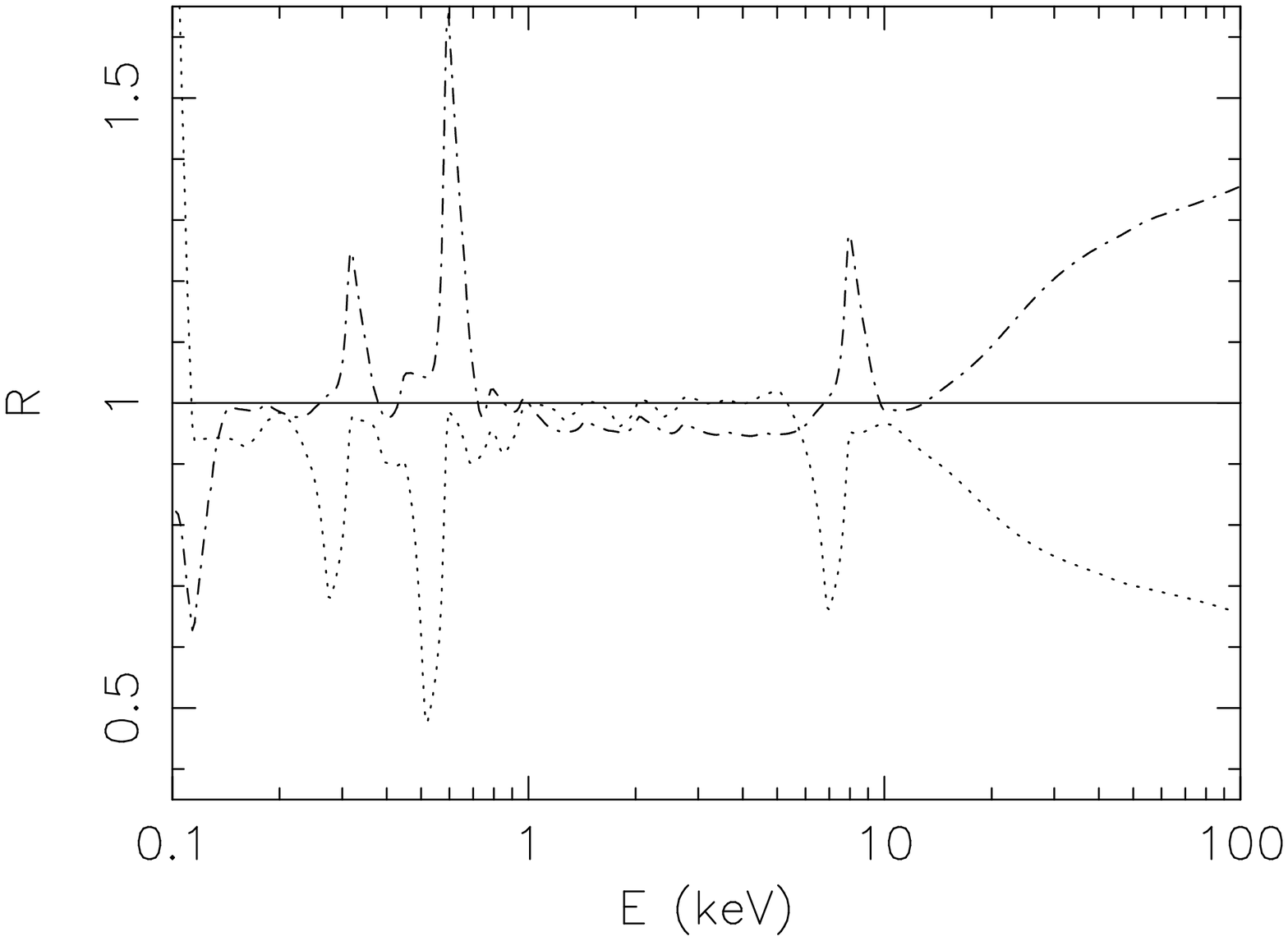}}  
\end{center}
\caption{Same as Fig.~\ref{rinplot} but for different disk inclination angles $i$. 
     Spectra for $i =1^{\circ}$, $45^{\circ}$ and $85^{\circ}$  
         (denoted by dotted, solid and dot-dashed lines respectively) are shown in the top panel, 
         and the corresponding quotient spectra, relative to the case $i = 45^{\circ}$  
           are in the bottom panel. } 
\label{incplot} 
\end{figure}

Figure~\ref{incplot} shows the spectra for disks with $i= 1^{\circ}$, $45^{\circ}$ and $85^{\circ}$.  
The intensity levels of the spectra of the disk with $i= 1^{\circ}$ and $45^{\circ}$ are similar. 
This is as expected for the following reasons.   
 At low inclination angles $i$  
     the dominant effect is the time dilation 
     (manifested as gravitational red-shift and transverse Doppler shift).   
Lensing is less important, 
     as it does not increase the projected area of the disk substantially when $i$ is small. 
The intensity is therefore still roughly proportional to $\cos i$,  
   and the cosines of $1^{\circ}$ and $45^{\circ}$ differ by a  factor less than unity.  
This proportionality no longer holds for larger inclination angles,  
   because when $i$ is sufficient large, gravitational lensing effects will be important.   
Nevertheless,    
   even when the enhancement due to gravitational lensing 
   and the additional contribution by the higher-order images are taken into account, 
   the total intensities for the edge-on disks are smaller than the face-on disks 
   provided that the index of the emissivity power law $\gamma \ge 2$ 
   (cf.\  the spectra of the case with $i= 85^{\circ}$ and the other two in Fig.~\ref{incplot}). 

Figure~\ref{incplot} also demonstrates that relativistic effects are more important as $i$ increases. 
The increases in the smearing of the edges can be seen 
    when comparing the spectra for $i= 1^{\circ}$, $45^{\circ}$ and $85^{\circ}$. 
Moreover, the spectra are shifted blue-ward when $i$ increases, 
   and the effects are more obvious in the quotient spectra. 
The increases in the blue shifts are due to the increase in the relativistic beaming/boosting 
   and the decrease in the role of time dilation, as $i$ increases.   
Note that the dip feature at energies above 10~keV that is seen in the quotient spectra 
   in Figures \ref{rinplot}, \ref{routplot} and \ref{emlawplot} is absent here.  

\section{Discussion}  

\subsection{General aspects} 

\begin{figure}[t] 
\begin{center} 
\vspace*{0.25cm} 
  \mbox{\epsfxsize=0.5\textwidth\epsfbox{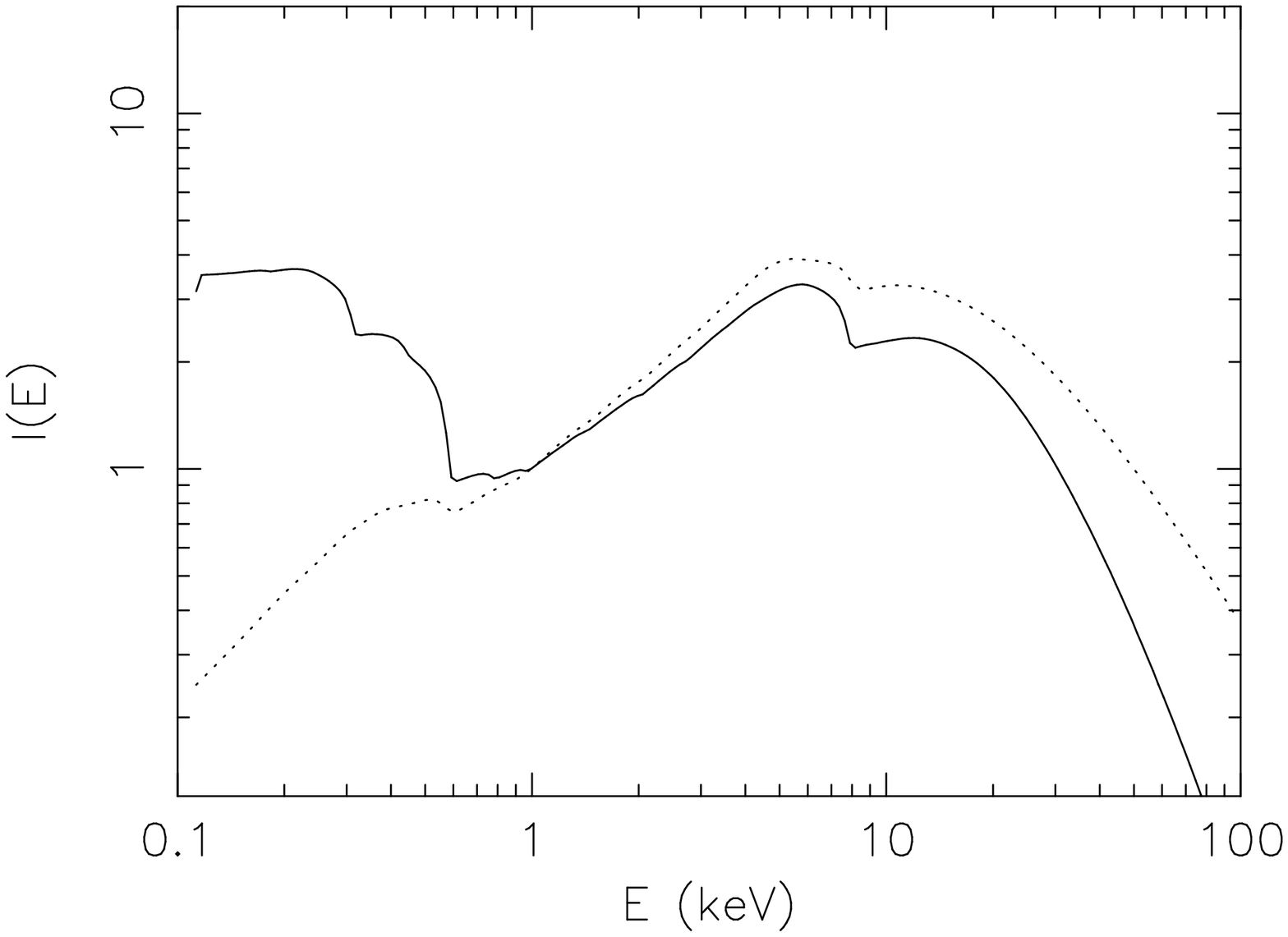}}    \\ 
  \mbox{\epsfxsize=0.5\textwidth\epsfbox{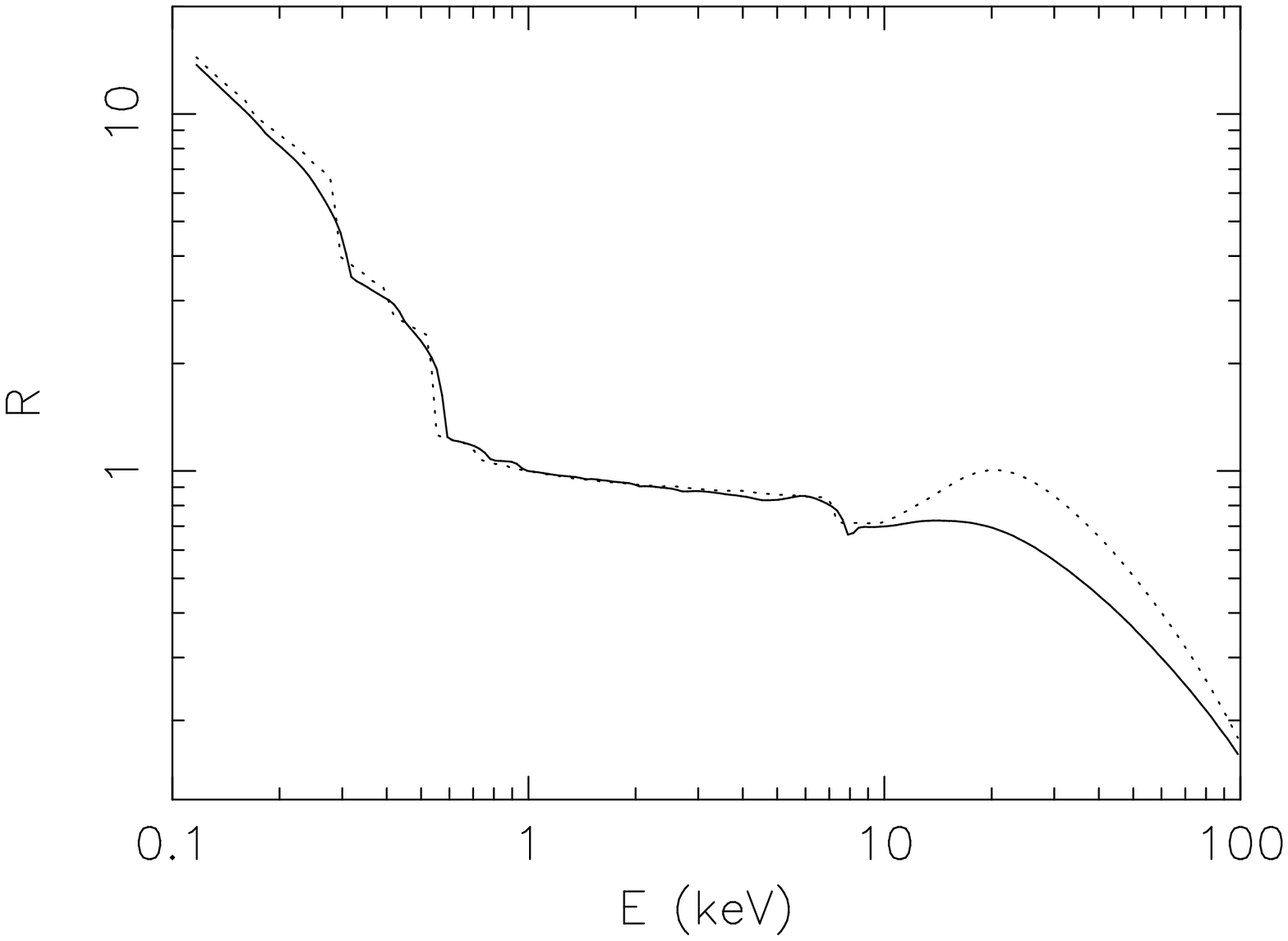}}  
\end{center}
\caption{
   A comparison of relativistic reflection spectra (top panel)  
      obtained by our convolution calculation (solid line) 
      and by the DYK04 calculation (dotted line).    
   A comparison of the ratio of the two spectra in the top panel (solid line) and 
      the ratio of the {\scshape pexrav} spectrum and the {\scshape hrefl} spectrum 
      (dotted line) (bottom panel). 
   The viewing inclinations in both case are 45$^\circ$ 
       and the spin-paramter of the black hole is 0.
   The other parameters are default as in Table 1. 
   The normalisations are such that $I(E) = 1$ and $R=1$ at $E=1$~keV.} 
\label{comp_1} 
\end{figure}

\begin{figure}[t] 
\begin{center} 
\vspace*{0.25cm} 
  \mbox{\epsfxsize=0.5\textwidth\epsfbox{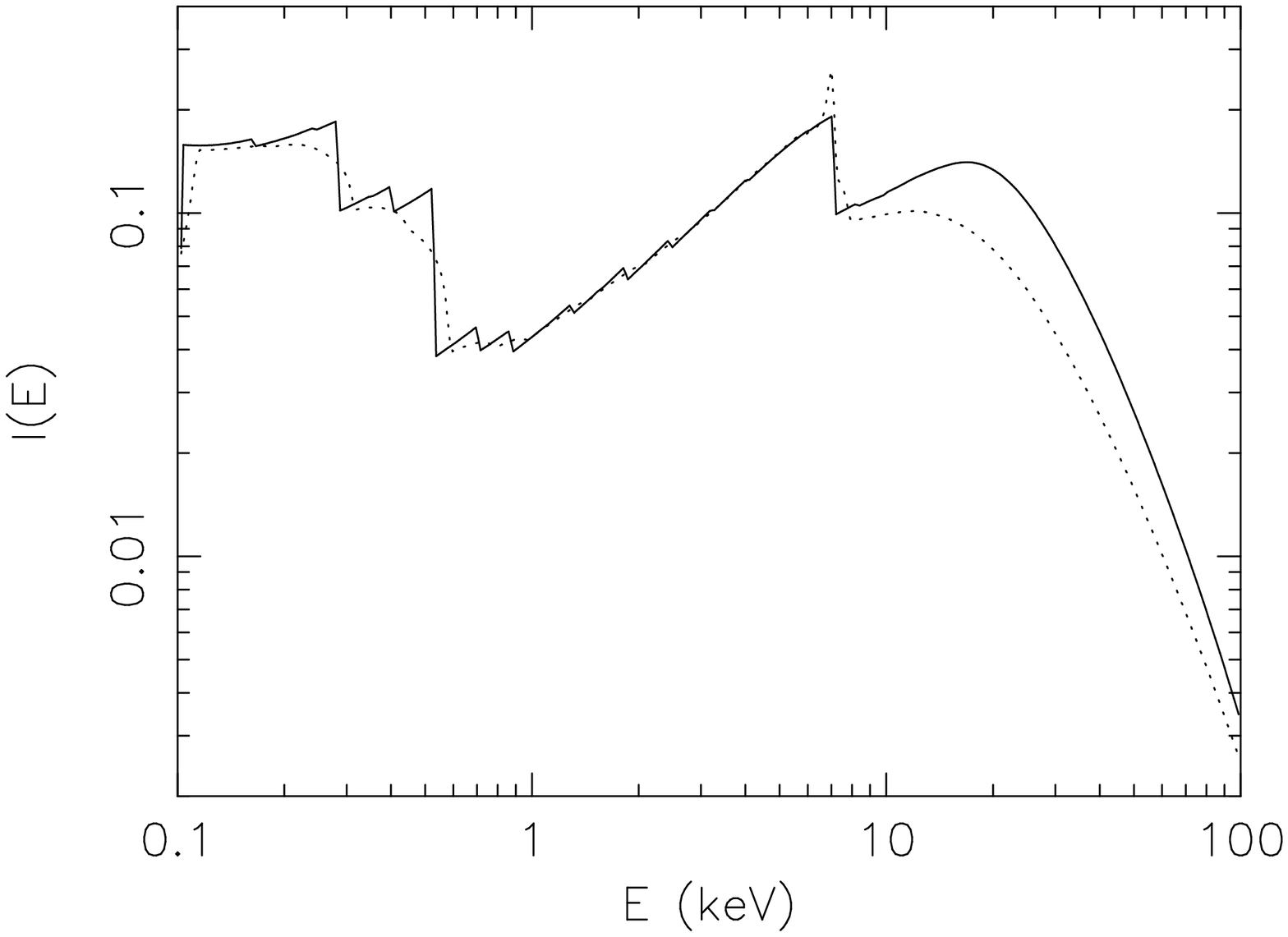}}    \\ 
  \mbox{\epsfxsize=0.5\textwidth\epsfbox{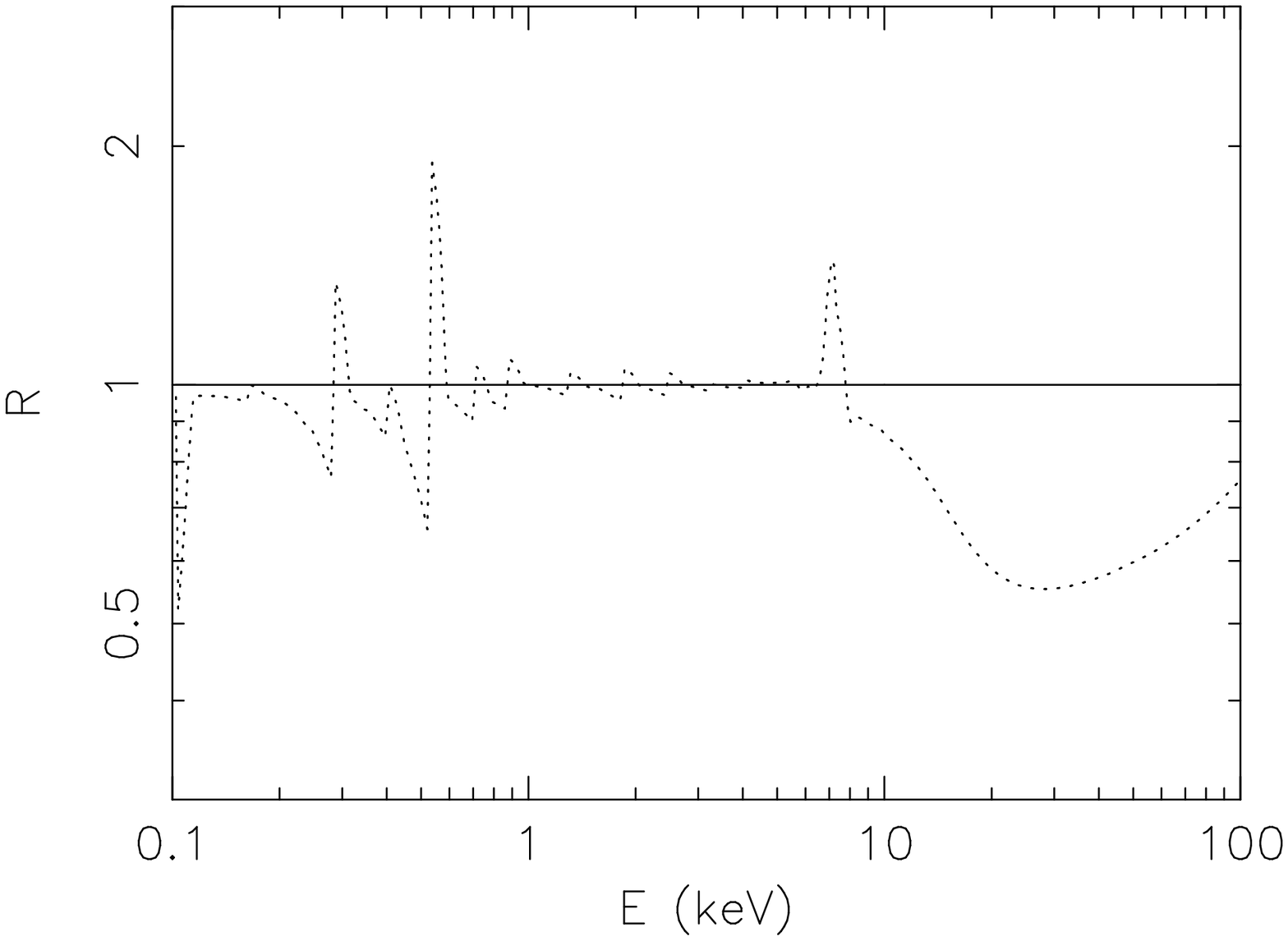}}  
\end{center}
\caption{Same as Fig.~\ref{shiftplot}, but a Fe K$\alpha$ line at 6.4~keV is added 
      to the relativistically modified reflection spectrum and the corresponding quotient spectrum.  } 
\label{lineplot} 
\end{figure}

Emissions from accretion disks around black holes 
   can be substantially modified by various relativistic effects.  
The energies of the emissions are shifted and the intensities are boosted      
   because of the high speeds of the accretion flows 
   and because of time dilation resulting from the transverse motion of the emitters 
   and the climbing of  the deep gravitational well by the photons.    
The emissions can also be modified by gravitational lensing and rotational frame dragging.  
These two effects are particularly important   
   if the emissions from the disk are anisotropic  
   and have strong spatial dependence.   
Some of the relativistic effects have already been demonstrated 
   in previous works on (isotropic) line profile calculations 
   (e.g.\  Fabian et al.\ 1989; Fuerst \& Wu 2004). 
Here, we show the relativistic effects in the spectra of reflection emission 
   of accretion disks, which is anisotropic and is subject to limb effects. 

As the reflection continuum is  angle dependent, 
   it is more affected by gravitational lensing and the rotational frame-dragging 
   than the isotropic line emission.   
In the absence of lensing, 
   the observable emissions from a flat, thin disk are expected to 
   have the same pitching angles as the disk viewing inclination angle $i$, 
   implying that the direction of the `observed' emissions from  a disk 
   with an almost edge-on viewing geometry  
   should be almost aligned with the tangent of the disk surface emitting elements.   
We have shown that because of lensing,   
   a substantial proportion of the observable emissions emerge at directions  
   aligned close to the normals of the emitting disk surface elements 
   even when the disk is viewed at an inclination angle $i = 85^\circ$.  
Therefore, 
   one cannot simply apply the energy shifts for the disk surface emitting elements 
   and convolve with the corresponding local reflection rest-frame spectra. 
The angle-dependence must be considered explicitly, as in this work,  
  when constructing the relativistic reflection model spectra.  

We have assessed how the distortions of the reflection spectra depend on the system parameters. 
Careful examinations of the spectra and the quotient spectra 
   reveal that certain effects are associated with some system parameters more than the others. 
The similarities between features in the quotient spectra 
   in Figures~\ref{rinplot},   \ref{routplot} and \ref{emlawplot} are not surprising, 
   as  the inner disk radius $r_{\rm in}$, outer disk radius $r_{\rm out}$ 
   and the emissivity power-law index $\gamma$ have similar roles.  
They determine the relative contributions of emission by the disk surface elements. 
Adjusting them changes the weights of emission from the inner disk regions 
   relative to the outer disk regions.    
In contrast, the smearing and energy shifts in the spectra in Figure~\ref{spinplot} are  
   different from those in the spectra in Figures~\ref{rinplot},   \ref{routplot} and \ref{emlawplot}. 
The difference is more obvious in the corresponding quotient spectra. 
In this work we have set the  inner disk radius to be the last stable particle orbit 
   which depends on the spin of the black hole only. 
The spin parameter $a$ therefore uniquely determines 
   how close the inner disk boundary can extend towards the black-hole event horizon. 
In the disk regions close to the black-hole event horizon, 
   time dilation is the dominant effect and it always red-shifts the energies of the emissions.  
The wedge-like features in the quotient spectra in  Figure~\ref{spinplot} 
   manifest the presence of large differential time dilation. 
The disk viewing inclination angle $i$ is, however, more effective 
   in causing energy shifting than smearing 
   (cf.\ the quotient spectra in Fig.~\ref{incplot} and \ref{spinplot}). 
This is due to the fact that 
   the projected line-of-sight velocity 
   is determined by $i$ (except at very small radii,  $r \ll  6 \ r_{\rm g}$), 
   and hence the main effect is relativistic Doppler shift.  
At large inclination angles, the `blue' emissions are strongly boosted.   
We can see blue-shifted features near the edges in the quotient spectrum 
    of the case $i = 85^\circ$ (Fig.~\ref{incplot}),    
   in contrast with that in the case of the black-hole spin, 
   where the features are always red-shifted.    
Moreover, the smearing is smaller, 
   because differential time dilation (due to gravitational red-shift)  
   across the strongly emitting regions is relatively small,  
   under the prescription that we have assumed for the inner disk boundary condition.   
   
\subsection{Comparison with other calculations}    

Calculations of relativistic reflection spectra have been carried out by various groups.  
Two representative calculations are those of 
  Martocchia, Karas \& Matt (2000) and Dovciak, Karas \& Yaqoob (2004) (hereafter DKY04).   
The numerical algorithm used in the convolution calculations of this study 
  is similar to that in DKY04.  
Our results  and those of DKY04 are in qualitative agreement. 
However, a more careful examination   
   reveals some subtle differences between the two calculations. 
These differences are sufficient to give rise to complications in spectral analysis 
   and may cause further dispute in the interpretations of relativistic features 
   in the X-ray spectra of AGN.  
 
We have used the {\scshape pexrav} model 
  to generate the input rest-frame reflection spectra, 
  while DYK04 used the {\scshape hrefl} model. 
In the $1-10$~keV range, the {\scshape hrefl} model, which is a reasonable approximation, 
  is a sensible choice 
  (given that it takes much longer computation time to use the {\scshape pexrav} model 
  to generate fine spectral grids).   
Outside the $1-10$~keV range, the two models differ very substantially, 
   and it is more appropriate to adopt the  {\scshape pexrav} model 
   for the input rest-frame spectral model.   
In Figure~\ref{comp_1} (top panel) we show 
  a relativistic reflection spectrum of an accretion disk  
  around a Schwarzschild black hole viewed at an inclination of 45$^\circ$ 
  obtained by our calculation  
  and a relativistic reflection spectrum of the DYK04 model 
  for the same parameters. 
The smeared and shifted metal edges below 0.5~keV  in our relativistic spectrum 
  are practically absent in the DKY04 spectrum.  
This is expected, as these edges are not obvious 
   the rest-frame {\scshape hrefl} model, 
   in contrast to the {\scshape pexrav} model,
   where the edge features at these energies are present.  
While these relativistic edges may be ignored 
   if we consider only narrow spectral regions near the Fe K$\alpha$ lines, 
   a proper treatment of the edges below 0.5~keV is essential 
   in the analysis of soft X-ray relativistic features,  
   such as those seen in the {\sl XMM-Newton} RGS spectra 
   (see e.g.\ Branduardi-Raymont et al.\ 2001).   

The different input reflection rest-frame spectra
  explain part of but not all the differences 
  between the relativistic reflection spectra obtained in this study and in DKY04. 
The relativistic kernel function in the convolution calculation also plays an important role. 
The kernel function used in DYK04 has taken into account 
  two main relativistic effects: the relativistic energy shifts and lensing 
  (a default for most relativistic calculations).  
Calculations in this work also consider these two effects.  
In addition, we use a full radiative transfer formulation    
   which takes account of anisotropy  and limb brightening/darkening of the emission. 
The anisotropic and limb effects are particularly important 
   for emission from the inner disk regions close to the central black hole, 
   where gravitational lensing and relativistic aberration are most severe. 
To illustrate these effects, we compare two quotient spectra:   
  the ratio of the {\scshape pexrav} model spectrum and the {\scshape hrefl} model spectrum 
  and the ratio of the corresponding relativistic reflection spectra.  
In the comparison we choose a moderate viewing inclination angle of 45$^\circ$, 
  so that the effects are not excessively exaggerated 
  as in the cases where the disk is viewed near edge-on. 
If the anisotropic and limb effects are unimportant, 
  the two spectral ratios should be similar at all energies.    
If they are not, the discrepancies in the two spectral ratios would be obvious.  
In the bottom panel of Figure~\ref{comp_1} 
  we show the ratio of the two relativistic reflection spectra presented in the top panel 
  and the ratio of the {\scshape pexrav} model spectrum and the {\scshape hrefl} model spectrum 
  viewed at 45$^\circ$. 
The two ratios are far from being identical.   
The difference between them is quite significant ($> 50$\%) at energies above 10~keV.  
In the analysis of observed AGN spectra,  
   such differences could cause large uncertainties  
   in determining the underlying continuum,  
   which affect substantially the fitting of the relativistic Fe K$\alpha$ line.      
The difference between the two ratios is less obvious at lower energies 
   but still visible. 
In particular, the location of the edges are different, 
   caused by weighting differences when anisotropy and limb effects are present.   
This comparison also demonstrates that 
   anisotropy and limb effects may distort the relativistically shifted edges, 
   mimicking and partially masking the gravitational effects.    

\subsection{The presence of fluorescent emission lines}   
 
So far we have considered only the reflection continua modified by relativistic effects  
   and compared them with the reflection rest-frame spectra. 
In more realistic situations, the reflection continuum is often accompanied by lines.  
The lines, which also originate from the inner accretion disk region, 
   are subject to relativistic effects. 
Figure~\ref{lineplot} shows a spectrum 
   consisting of a Fe K$\alpha$ line and a reflection continuum component 
   and its corresponding quotient spectrum.   
(The line is not included in the rest-frame model spectrum to illustrate 
    the combined effects of shift/smearing of the edges and the presence of a relativistic line.) 
The inclusion of the line causes complications.  
This simple example demonstrates that 
    the relativistic shifted red wing of the Fe line can `fill up'  its corresponding reflection edge. 
   
Fluorescent X-ray emission lines, in particular, the Fe K$\alpha$ line,  
  are potentially very powerful tools to diagnose 
  the space-time distortion and various dynamical relativistic effects  
  in flows near the event horizons of accreting black holes in AGN.  
In principle, we can derive the size and the physical condition 
  of the inner rim of an accretion disk 
  from the relativistic shift and broadening of the lines.  
This information allows us to deduce various system parameters 
  such as the black-hole spin.  
(For a review of this subject, see e.g.\  Fabian et al.\ 2000; Miller 2007.)      
However, complications may arise when it is put into practice.  
Firstly, there are parameter degeneracies.   
Also, there are uncertainties in accretion disk modeling. 
For instance,  
  if the accretion disk is geometrically thick (see e.g.\ Fuerst \& Wu 2007) 
  or if there are magnetic-field controlled outflows 
  or other complicated "microscopic processes" in the inner disk 
  (see e.g.\ Krolik \& Hawley 2002; Yuan et al.\ 2008),    
  we would not have a well defined one-to-one relation 
  between the last innermost stable orbit for the (free-falling) particles 
  and the inner radius of the accretion disk as in the case of geometrically thin Keplerian disks.   
In these situations, knowing the inner disk radius  
  does not automatically give us the black-hole spin.    
Secondly, how reliably we extract the black-hole parameters from the lines   
  depends on how accurately we determine the relativistic distortion of the  line profile.  
From what we have shown above, 
  we can see that it is a non-trivial task 
   to separate the lines from the reflection continuum in an observed X-ray spectrum  
   (especially in noisy low-resolution data) 
   when both the components are severely shifted and smeared by various relativistic effects.   
Subtracting the continuum separately before fitting the lines will certainly lead 
  to bias and inconsistencies.  
To improve the extraction of black-hole parameters using fits to observed X-ray spectra, 
  simultaneous modeling of relativistically distorted continuum and emission lines is needed.  
   
\section{Summary}    
    
We have convolved angle-dependent rest-frame reflection spectra of accretion disks 
   with relativistic ray-tracing radiative transfer calculations 
   to generate relativistically modified reflection model spectra of accretion disks in AGN. 
We have shown that the resultant spectra are significantly modified, 
    with edge smearing and energy shifting. 
The degrees of edge smearing and energy shifting depend on 
   the accretion disk parameters as well as the black-hole spin. 
The smearing of the edges has strong dependence 
   on the accretion disk outer and inner radii, the black-hole spin, 
   and the disk emissivity profile. 
Energy shifting is, however, more affected by the accretion disk viewing inclination angle. 
Anisotropy and limb effects can sometimes  mimick and partially mask the gravitational effects.  
These factors complicate the interpretation of X-ray spectral observations of AGN, 
  as it is not straightforward to disentangle the effects 
  on the lines, the edges and the underlying multi-component continuum 
  of severely relativistically modified AGN spectra.   
Despite these, 
  simultaneously modeling the relativistic effects 
  on the continuum and the emission lines in spectral analyses 
  would reduce bias uncertainty in determine the relativistic line profiles 
  and hence the black-hole parameters.  

\begin{acknowledgements} 
This work was supported by the Nuffield Foundation through the Research Bursary program. 
\end{acknowledgements}

\label{lastpage}

\end{document}